\documentclass[journal]{IEEEtran}
\ifCLASSINFOpdf
\else
\fi

\usepackage [latin1]{inputenc}
\usepackage{amsmath,amsfonts,amssymb}
\usepackage{algorithm}
\usepackage{algorithmic} 
\usepackage{booktabs}
\usepackage{multirow}
\usepackage{bigstrut}
\usepackage{longtable}
\usepackage{graphicx}
\usepackage{epstopdf}
\usepackage[square, comma, sort&compress, numbers]{natbib} 
\usepackage[numbers]{natbib}
\usepackage{gensymb}
\usepackage{enumerate}
\usepackage{bibspacing}
\usepackage{url}
  \usepackage[usenames, dvipsnames]{color}

\newcommand{\cred}[1]{\textcolor{red}{#1}}

\usepackage[pdftex]{hyperref} 
\hypersetup{colorlinks,       
            linkcolor=blue,  
            anchorcolor=black,
            citecolor=blue,  
            filecolor=black,  
            menucolor=black,  
            pagecolor=black,  
            urlcolor=red,   
	          bookmarks=true,         
	          bookmarksopen=false,    
	          bookmarksnumbered=true, 
	       }
\usepackage{cleveref}
\crefname{table}{Tab.}{Tabs.}

\begin{document}
%
\title{Hyperspectral Image Denoising and Anomaly Detection Based on Low-rank and Sparse Representations}
%
%
%

\author{Lina~Zhuang,~\IEEEmembership{Member,~IEEE,}
        Lianru~Gao,~\IEEEmembership{Senior~Member,~IEEE,}
        Bing Zhang,~\IEEEmembership{Fellow,~IEEE,}
        Xiyou Fu, ~
\IEEEmembership{Member,~IEEE,}
        and~Jos\'{e}~M.~Bioucas-Dias,~\IEEEmembership{Fellow,~IEEE}
\thanks{This work 
was supported in  part  by  the  Hong  Kong  Baptist  University  Start-up  under  Grant 21.4551.162562, and supported by the National Natural Science Foundation of China under Grant 42001287 and Grant 41722108. (\textit{Corresponding author: Lianru Gao.})}
\thanks{L. Zhuang is with the Department of Mathematics, Hong Kong BaptistUniversity, Hong Kong (e-mail: linazhuang@qq.com).}
\thanks{L.  Gao  is  with  the  Key  Laboratory  of  Digital  Earth  Science, Aerospace Information Research Institute, Chinese Academy of Sciences, Beijing 100094, China.
}
\thanks{B. Zhang is with the Key Laboratory of Digital Earth Science, Aerospace Information Research Institute, Chinese Academy of Sciences, Beijing 100094, China, and also with the College of Resources and Environment, University of Chinese Academy of Sciences, Beijing 100049, China (e-mail: zb@radi.ac.cn).}
\thanks{X. Fu is with the Guangdong Laboratory of Artificial Intelligence and Digital Economy (SZ), Shenzhen University, Shenzhen 518060, China, and also with the College of Computer Science and Software Engineering, Shenzhen University, Shenzhen 518060, China (e-mail: fuxiyou@qq.com).}
\thanks{J. M. Bioucas is with Instituto de Telecomunica\c{c}\~{o}es, Instituto Superior T\'{e}cnico, Universidade de Lisboa, Lisbon 1049-001,  Portugal (e-mail: bioucas@lx.it.pt).}
}

%



\maketitle

\begin{abstract}
Hyperspectral imaging measures the amount of electromagnetic energy across the instantaneous field of view at a very high resolution in hundreds or thousands of spectral channels. This enables objects to be detected and the identification of materials that have subtle differences between them. However, the increase in spectral resolution often means that there is a decrease in the number of photons received in each channel, which means that the noise linked to the image formation process is greater. This degradation limits the quality of the extracted information and its potential applications. Thus, denoising is a fundamental problem in hyperspectral image (HSI) processing. As images of natural scenes with highly correlated spectral channels, HSIs are characterized by a high level of self-similarity and can be well approximated by low-rank representations. These characteristics underlie the state-of-the-art methods used in HSI denoising. However, where there are rarely occurring pixel types, the denoising performance of these methods is not optimal, and the subsequent detection of these pixels may be compromised. To address these hurdles, in this paper, we introduce {\bf RhyDe} ({\bf R}obust {\bf hy}perspectral {\bf De}noising), a powerful HSI denoiser, which implements explicit low-rank representation, promotes self-similarity, and, by using a form of collaborative sparsity, preserves rare pixels. The denoising and detection effectiveness of the proposed robust HSI denoiser is illustrated using semi-real and real data.
\cred{A MATLAB demo of this work is available
at \url{https://github.com/LinaZhuang} for the
sake of reproducibility.}
\end{abstract}


%
\IEEEpeerreviewmaketitle

\section{Introduction}

Hyperspectral imaging consists of the measurement of electromagnetic energy across an instantaneous field of view in hundreds or thousands of spectral channels with a higher spectral resolution than multispectral/RGB cameras. Whereas the human eye perceives the visible wavelengths of light (namely, red, green, and blue), spectrometers cover a wider range of wavelengths (beyond the visible) and divide the spectrum into many more bands with a fine wavelength resolution. The characteristics of high spectral resolution imagery enable the precise identification of materials via spectroscopic analysis. However, the amount of noise in the measurements often precludes the widespread use of hyperspectral images (HSIs) in applications that require precise material identification (e.g., in precision farming).

As a result of recent developments, self-similarity and low-rank-based image denoising can be considered state-of-the-art in HSI denoising \cite{FastHyDe,he2015hyperspectral,he2019non}. As natural images, HSIs are self-similar. This means that they  contain many similar patches  at different locations or scales \cite{buades2005non,teodoro2015single,BM3D}. This characteristic has been recently
exploited by patch-based image restoration methods:
BM4D \cite{BM4D}, VBM4D \cite{VBM4D}, and MSPCA-BM3D \cite{danielyan2010denoising} 
use collaborative filtering in groups of 3-D patches extracted from volumetric data, videos, and multispectral data, respectively.
DHOSVD \cite{HOSVD} applies a hard threshold filter  to    coefficients of higher order SVD of similar patches.   
Similar ideas have been pursued in HSI denoising:
3-D nonlocal sparse denoising \cite{Qian2013nonlocal,qian2013NLM,Qian2013NLtensor,LiaoTensor} uses similar 3-D blocks for sparse coding; `PCA+BM4D' \cite{pca+vbm4d} attempts to decompose the observed HSI into signal and noise using PCA and applies BM4D filtering to low-energy components corresponding to noise; and FastHyDe \cite{FastHyDe}  exploits the  self-similarity  of the HSI representation coefficients in suitable subspaces. 

Besides exploiting the high spatial correlation of HSIs by using patch-based regularization, the high spectral correlation has been   investigated mainly by means of low-rank representation and total variation minimization.
The high spectral correlation results in a low-rank structure for the HSI, which allows extremely compact and sparse representations within suitable frames. We refer readers to these works: the noise adjusted iterative low-rank matrix approximation (NAILRMA) \cite{he2015hyperspectral}, the structured tensor TV-based regularization \cite{SSAHTV,ZhaoTV}, or the adaptive spectrum-weighted sparse Bayesian dictionary learning
method (ABPFA) \cite{CS_spectra}. The very high spectral resolution of hyperspectral imagery leads to the acquisition of piece-wise smooth spectral vectors, which can be characterized by minimizing the total variation of spectral vectors \cite{SSAHTV,ZhaoTV}. For example, SSTV \cite{aggarwal2016hyperspectral}  removes hyperspectral mixed noise by utilizing the total 1-D variation along the spectral dimension and the total 2-D variation along the spatial dimension.

The existence of a large number of bands usually leads  to the high computational complexity for most HSI denoisers. 
In fact, this computational burden can be greatly alleviated by taking advantage of the high correlation between spectral channels.
For instance,  instead of denoising the original HSI, FastHyDe \cite{FastHyDe} and GLF \cite{GLF}   formulate the denoising with respect to the representation coefficients of the HSI subspace by exploiting the fact that  hyperspectral vectors exist, to a very good approximation, in subspaces that have a small number of dimensions compared with the number of bands \cite{overview}. 
This subspace may be accurately inferred from observations of HSIs corrupted by additive Gaussian noise \cite{Hysime}. 
This is  the strategy followed by FastHyDe \cite{FastHyDe}, GLF \cite{GLF}, and NG-meet \cite{he2019non},which, to the best of our knowledge, can be considered the state-of the-art methods for attacking  Gaussian noise in HSIs; they are also faster than their competitors.

However, the presence of 'rare pixels' may degrade the denoising performance and preclude the future detection of those pixels. The term 'rare pixels' refer to pixels corresponding to surface materials that occur only rarely in the imagery and thus have spectra that differ significantly from those of the majority of the pixels (often called the background). Due to the low probability of their occurrence, rare pixels may contribute little to the dictionaries/basis learned directly from observed images. The spectra of rare pixels may be corrupted by denoising since rare spectra are not well approximated by low-rank and sparse representations. Rare pixels, however, may contain important information that can be used in subsequent applications - for example, in detection processes.

The detection of rare pixels, often termed the  anomaly detection problem \cite{overviewAnomaly}, has been the object of considerable research efforts, some of which has been devoted to  the development of accurate models for the background. Representative models for the background include  Gaussian density \cite{RX,OSP-based}, mixtures of Gaussian density \cite{Gaussian-mixtureModel}, and (sparse) linear mixing models \cite{NRS_WeiLi,BSJSBD,8330051,YingQu}. Sparsity is exploited in the linear mixing models based on the assumption that each pixel in the background can be well represented sparsely using a background dictionary, whereas anomalies can not be. In this case, how to find a dictionary that completely represents the background materials but excludes anomalies is of crucial importance. In anomaly detection based on low-rank and sparse representation (LRASR) \cite{ZWu}, this background dictionary is selected from the original image pixels. All of the pixels are separated into different clusters using K-means and pixels that are closer to the mean value of each cluster (in the sense of the Euclidean distance) are selected as background pixels. In collaborative representation based detection (CRD) \cite{NRS_WeiLi}, the background dictionary is adaptive to pixels and is composed of pixels that are spatial neighbors. The main objective in anomaly detection is to classify a pixel as either background or anomaly if it does not fit the background model. Given the usual complexity of the land cover in remote sensing imagery, this classification problem is often quite challenging. 

Recently, besides the development of background models, the structure of anomalous pixels has also been investigated. Anomalies are often spatially sparse, a situation which is expressed as  group sparse regularization in LRASR \cite{ZWu}.
Orthogonal subspace projection (OSP)-based approaches \cite{OSP_CIChang,chang2005orthogonal} improve the detection of signals by projecting data onto a subspace that is orthogonal to the one spanned by the background signatures. By doing this, the background components are suppressed. OSP is usually regarded as a preprocessing step used before the application of other detectors \cite{OSP-based,overviewAnomaly}. 

The aim of the present study was to endow our previously developed FastHyDe denoiser \cite{FastHyDe} with the ability to preserve rare pixels, which might be important targets in subsequent applications. The proposed method takes full advantage of the structural sparsity linked to the usual highly sparse distribution of rare pixels in the HSI spatial domain. In addition, we also introduce an anomaly detection method, which is a spin-off from the denoiser algorithm and is competitive with the state-of-the-art anomaly detection methods.
 
We give the proposed denoising and anomaly detection framework 
the name {\bf R}obust {\bf hy}perspectral {\bf De}noising ({\bf RhyDe}).
This work is an extension of the material published in \cite{RhyDe_cof}.
The new material includes the following. a) RhyDe is introduced and described in more detail. The prior knowledge used in RhyDe, including the low rank and self-similarity of HSIs and the structural sparsity of rare pixels, is introduced in detail. b) An exhaustive array of experiments and comparisons is carried out. The impact of the denoising step on anomaly detection tasks is tested using two simulated HSIs and two real HSIs.  

The paper is organized as follows. Section 2 formally introduces RhyDe along with a denoising approach for the preservation of rare pixels and a derived anomaly detector. Sections 3 and 4 present the experimental results, including comparisons with those obtained using state-of-the-art methods. Section 5 concludes the paper.

\section{Problem Formulation}

\subsection{Observation Model}
 
 It is assumed that the dataset contains a small number of spectral/spatial outliers in unknown positions. These outliers are usually rare pixels with a low probability of occurrence. In a similar way to robust PCA \cite{robustPCA} and to the formulations in \cite{he2015hyperspectral,FastHyDe}, we adopt the observation model:
 \begin{equation}
 \label{eq:ob_m}
 {\bf Y} = {\bf X} + {\bf S} + {\bf N},
 \end{equation}
which assumes that the observation data matrix ${\bf Y} \in {\mathbb R}^{n_b \times n}$, with $n_b$ rows (spectral bands) and $n$ columns (spatial pixels)
is the sum of   matrix ${\bf X}  \in {\mathbb R}^{n_b \times n}$, representing the signal component, ${\bf S}  \in {\mathbb R}^{n_b \times n}$, representing the rare pixels, and ${\bf N}  \in {\mathbb R}^{n_b \times n}$, which represents
the noise and modeling errors. 

Given  ${\bf Y}$, our objective  is the estimation of ${\bf X}$ and $\bf S$, by exploiting the inherent structure of ${\bf X}$ and $\bf S$. The high spectral correlation of HSIs allows a low-rank approximation of matrix ${\bf X}$. Furthermore, the underlying image ${\bf X}$ is self-similar since the imaged objects consist of natural scenes on the Earth's surface. The matrix ${\bf S}$ is assumed to be column-wise sparse due to the fact that  rare pixels are sparsely distributed in the spatial domain. 
Below, we discuss the structure of ${\bf X}$ and ${\bf S}$   in detail. 

 \begin{figure*}[htbp]
\includegraphics[width=17cm]{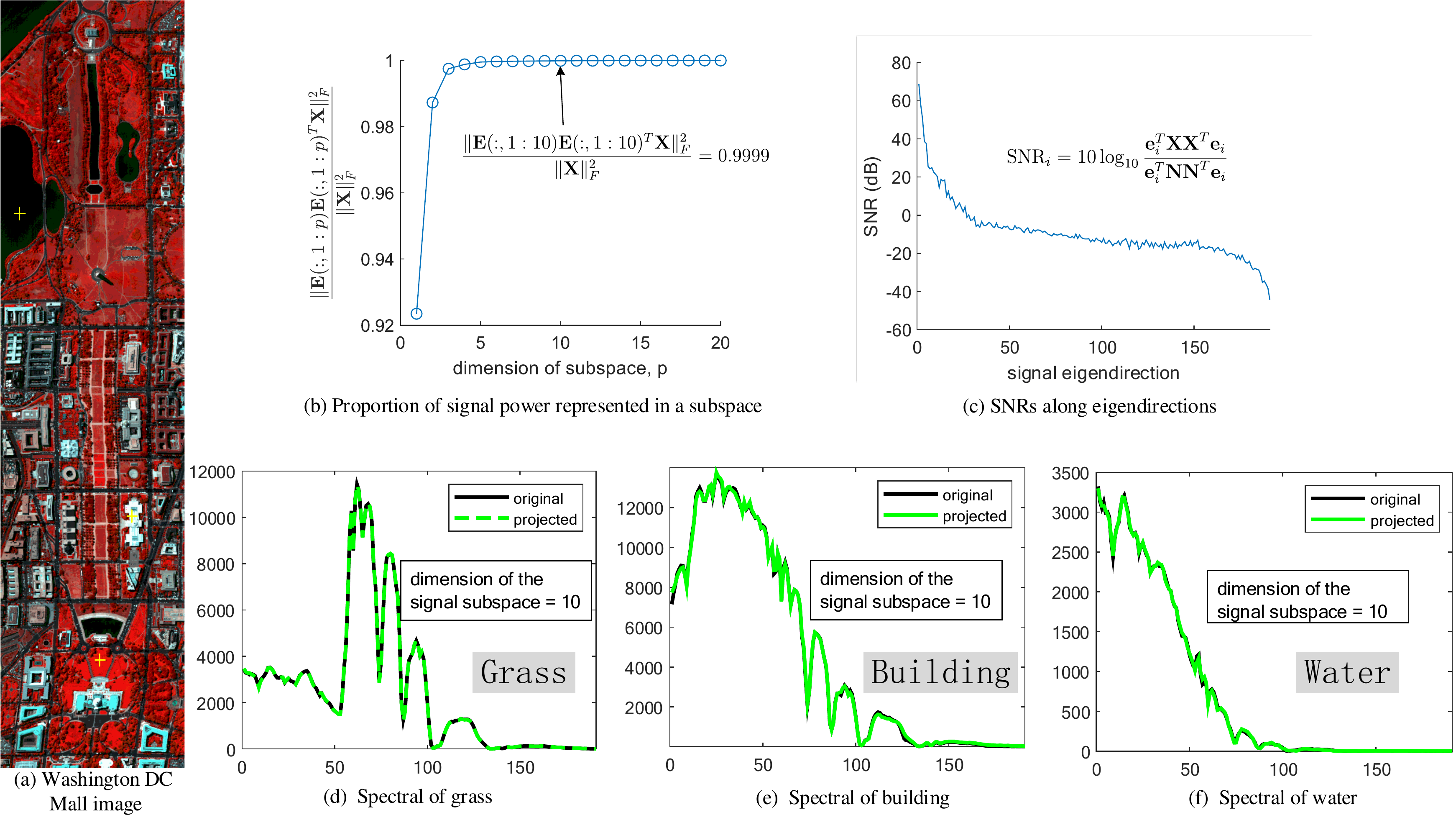}
\caption{Subspace representation of spectral vectors in Washington DC Mall image of size $1280\times 307 \times 191$.}
 \label{fig:subspace_ex}
\end{figure*}

\subsection{Structure of Signal Component ${\bf X}$ and Outlier Component ${\bf S}$}
\subsubsection{Subspace representation of ${\bf X}$}

Given the very high spectral correlation, we assume that  
 the columns (spectral vectors) of matrix ${\bf X}$ 
 live in a low-dimensional subspace that may be estimated from the
observed data $\bf Y$ to a good approximation \cite{overview,FastHyDe}.
 Am example  is given in Fig. \ref{fig:subspace_ex} to  illustrate the effectiveness of the subspace representation of  ${\bf X}$. 
The spectral vectors of the Washington DC Mall   image are projected onto a subspace with dimension 10 estimated by the HySime method \cite{Hysime}.  
We can see from Fig. \ref{fig:subspace_ex}-(d-f) that the  projected spectra of the classes grass, building, and water are close to their original spectra.
The spectral vectors are well represented in a subspace of dimension 10 due to the fact that $99.99\%$ of the signal power remains in the subspace (see Fig. \ref{fig:subspace_ex}-(b)). 
 The orthogonal subspace is an extremely compact representation of the image  since the signal is highly concentrated in the first few eigendirections (corresponding to the high SNRs shown in Fig. \ref{fig:subspace_ex}-(c)), and the eigendirections corresponding to   near-zero singular values mainly contain noise (corresponding to the very low SNRs shown in Fig. \ref{fig:subspace_ex}-(c)).

Based on the subspace representation, we write 
\begin{equation}
 \mathbf{X=EZ},
\end{equation}
with ${\bf E} = [{\bf e}_1, {\bf e}_2, \dots, {\bf e}_p] \in\mathbb{R}^{n_b\times p}$ ($p\ll n_b$)   being an orthogonal basis for the signal subspace and ${\bf Z} \in\mathbb{R}^{p\times n}$ containing the representation coefficients for ${\bf X}$ with respect to ${\bf E}$. 
We  will  use the  term \textit{eigen-images} for the  images  associated  with  the rows of ${\bf Z}$.

The signal subspace ${\bf E}$ can be estimated from the observation ${\bf Y}$ by simply carrying out
 singular value   decomposition  (SVD)  of ${\bf Y}$  in  cases where the noise is independent and identically distributed (i.i.d.) because the noise increases  singular values in the direction of each eigenvector  uniformly and does not change the order of these singular   values (and thus does not change the estimation of the subspace).
%
The signal subspace is approximately spanned by   $p$ singular vectors of ${\bf Y}$ corresponding to the top $p$ singular values:
 \begin{equation}
 \label{eq:learnE}
 {\bf E} = {\bf U}(:,1:p),
 \end{equation}
where ${\bf U} \in \mathbb{R}^{n_b \times n_b}$ is an orthogonal matrix and   $\{ {\bf U}, {\bf \Sigma}, {\bf V} \} = \text{SVD}({\bf Y})$ with the   singular values ordered by non-increasing magnitude.
The dimension of subspace, $p$, can be estimated by some subspace identification methods, such as HySime \cite{Hysime} and HFC \cite{harsanyi1993determining}.

Generally, there are  several ways to promote low-rankness of the image: a) we can directly impose a rank regularization on the image matrix, that is  rank(${\bf X}$);
b) We can minimize the nuclear norm of ${\bf X}$, which is a relaxed  rank constraint;
and
c) the rank constraint
can be equivalently transformed since the matrix can be factorized into
two smaller-sized matrices, i.e., ${\bf X} = {\bf EZ}$.
In HSI denoising problem, we use the third way, which brings some benefits:
The rank of image matrix is explicitly determined by the sizes of two smaller-sized matrices, instead of being implicitly affected by  parameters of regularizations in a) and b). Users can set the rank of   matrix ${\bf X}$ directly by changing the size of matrix ${\bf E}$, which is corresponding to the dimension of subspace, $p$. Furthermore, as given in \eqref{eq:learnE}, matrix ${\bf E}$ can be leaned from noisy observations with very good approximation. Instead of estimating ${\bf X} \in \mathbb{R}^{n_b \times n}$, we only need to estimate ${\bf Z} \in \mathbb{R}^{p \times n}$ ($p\ll n_b$), meaning the size of unknown variables has been reduced greatly. 
Therefore, an implementation of explicit low-rank representation of the HSI
improves  the  conditioning  of  the  denoising problem.

\subsubsection{Self-similarity of ${\bf Z}$}

\begin{figure}[htbp]
\includegraphics[scale=0.4]{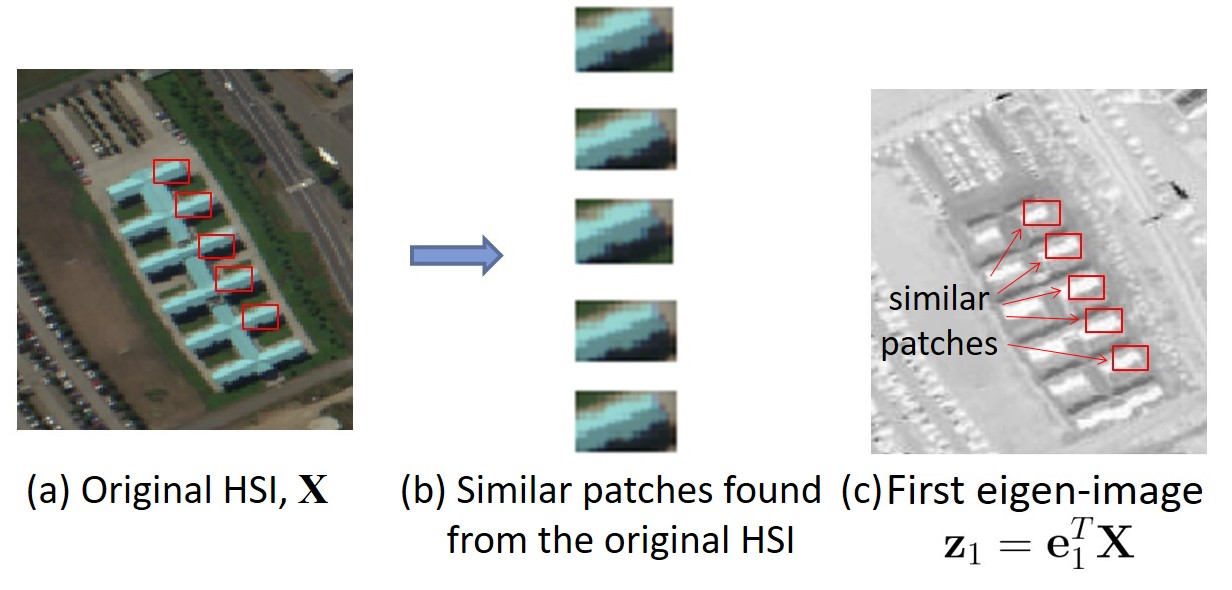} 
\caption{Self-similarity property of a sub-scene of Pavia University image and its subspace coefficients}
\label{fig:selfsimilarity}
\end{figure}

HSIs are self-similar, meaning that when we observe an image from small patches perspective,  
  similar   patches can be found at different locations and scales in the spatial domain (see Fig. \ref{fig:selfsimilarity}-(a,b)). 
These non-local similar patches are grouped together, and these groups have a much higher sparsity than an individual local patch. The high sparsity allows us to remove noise efficiently by using sparse representations of the grouped patches. This is the main idea behind the non-local patch-based image denoisers (such as BM3D \cite{BM3D} and WNNM \cite{gu2014weighted}). Also, the exploitation of self-similarity underlies the state-of-the-art methods that are applied to  inverse problems in imaging \cite{8705392,HyDemosaicing}.

Not only is the original HSI self-similar, but also its corresponding eigen-images are self-similar (see  Fig. \ref{fig:selfsimilarity}-(c)).
Each eigen-image, which is associated with each row of ${\bf Z}$, is a linear combination of the bands of the hyperspectral image; i.e., ${\bf z}_i  = {\bf e}_i^T {\bf X}, (i = 1, \dots, p)$. 
In all the bands, electromagnetic energy is acquired from exactly the same objects, which have the same spatial structure; thus all the eigen-images contain the same spatial structure.
The high spatial correlation of eigen-images enables us to exploit the 
 self-similarity of ${\bf Z}$   as a type of regularization in this work.

\subsubsection{Column-wise sparsity of ${\bf S}$}

\begin{figure}[htbp]

\includegraphics[scale=0.4]{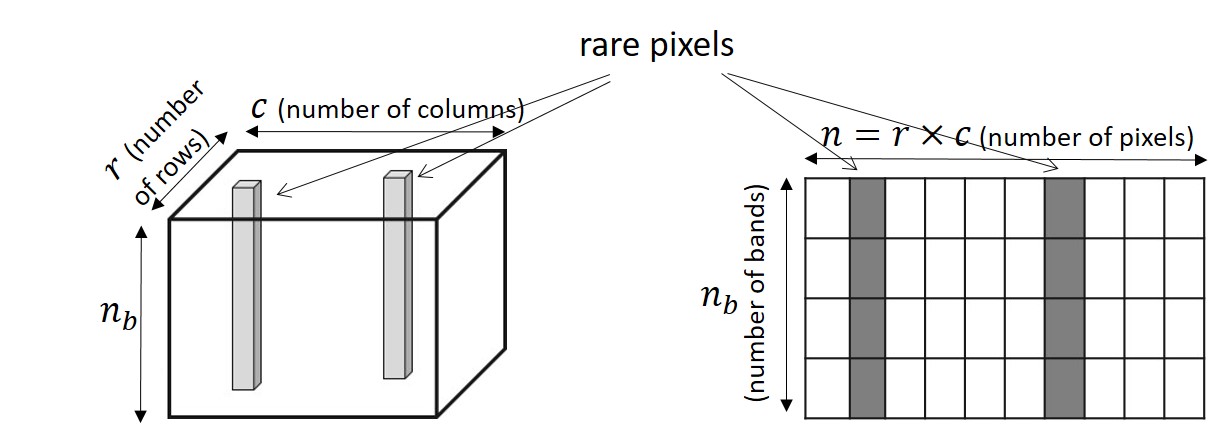}
\begin{tabular}{lc}
\footnotesize{  (a) Spatial sparse rare pixels in HSI}
   &  \footnotesize{  (b) Column-wise sparsity} \\
& \footnotesize{Mixed norm penalty:}\\
& \footnotesize{$\|{\bf S}^T\|_{2,1}= \sum_{i=1}^n \|{\bf s}_i \|_2$ }\\
&\footnotesize{(${\bf s}_i$ denotes $i$-th column of ${\bf S}$)}
\end{tabular}
\caption{Column-wise sparsity of matrix ${\bf S}$ representing rare pixels}
\label{fig:priorS}
\end{figure}

As `rare pixels' means pixels that contain materials that occur rarely in the image, these materials are sparsely distributed in the spatial domain (see Fig. \ref{fig:priorS}-(a)). When utilizing low-rank and sparse representations of HSIs, the spectral vectors of rare pixels may not be represented well in the frames/basis as these are learned from observations, and rare pixels, with their low probability, contribute little to the image signal and thus have little effect on the estimation of the frames or basis. Although their  probability may be low, rare pixels may contain important information that can be used in subsequent applications - for example, in target detection.

To preserve the rare-pixel information in the subspace representation framework, we modeled the rare pixels (which are not represented well in the subspace) as an additive term ${\bf S}$ in \eqref{eq:ob_m}. As the distribution of rare pixels  in the spatial domain is usually highly sparse,  matrix ${\bf S}$ should be column-wise sparse (see Fig. \ref{fig:priorS}-(b)) and can be characterized by a mixed norm.

As stated above, three characteristics of HSIs are exploited in this paper. 
\begin{itemize}
\item[a)] The spectral vectors of HSIs can be well represented in   low dimensional subspaces
\item[b)] Images of the subspace representation coefficients, called eigen-images in this paper, are self-similar and thus suitable for denoising using non-local patch-based methods.
\item[c)] Rare pixels (anomalies) are often spatially sparse.
\end{itemize} 

We first simplified the denoising problem by assuming that the noise was additive Gaussian and i.i.d. Although the noise in real HSIs rarely conforms to this assumption, this simplification leads to a better understanding of how signal and noise are changed moving from the original high-dimensional space to the projected low-dimensional subspace. Later, we consider additive Gaussian non-i.i.d. noise, which  corresponds more closely to the situation encountered in reality.

\subsection{Denoiser for dealing with Gaussian i.i.d. Noise}
\label{sec:iid}

\subsubsection{Ojective function}

Based on the above rationale and the assumption of Gaussian i.i.d. noise, we propose  to estimate the matrix $\bf Z$ and the sparse matrix $\bf S$ , which represents the outliers, by solving the optimization 
\begin{equation}
\label{eq:2unc}
\{ \widehat{{\bf Z}},\widehat{{\bf S}} \} \in \arg \min_{{\bf Z}, {\bf S}} \frac{1}{2} || {\bf EZ} + {\bf S} - {\bf Y} ||_F^2 + \lambda_1 \phi ({\bf Z}) + \lambda_2 || {\bf S}^T||_{2,1},
\end{equation}
where $||{\bf X}||_F^2 = \text{trace}({\bf XX}^T)$ is the Frobenius norm of ${\bf X}$. The first term on the right-hand side represents the data fidelity. 
The second term is a regularization expressing prior information tailored to self-similar images \cite{buades2005non,BM3D,LRCF,FastHyDe}. 
Note that here we do not give an explicit definition of the function $\phi(\cdot)$, because we use plug-and-play prior \cite{Plug-and-Play} for ${\bf Z}$. 
The central idea in plug-and-play is   to use directly an off-the-shelf denoiser (such as BM3D \cite{BM3D} and WNNM \cite{gu2014weighted}) conceived to enforce self-similarity, instead of investing efforts in tailoring regularizers promoting self-similar images and then computing its proximity operators. 
Therefore the definition of $\phi(\cdot)$ depends on the denoisers plugged into the algorithm (see Section~\ref{sec:ppp} for further details).
The third term is a regularization of the sparse matrix ${\bf S}$. 
The mixed  $\ell_{2,1}$ norm of ${\bf S}^T$, given by $\|{\bf S}^T\|_{2,1}= \sum_{i=1}^n \|{\bf s}_i \|_2$ (${\bf s}_i$ denotes $i$-th column of ${\bf S}$),   promotes column-wise sparsity   of ${\bf S}$ (see, e.g., \cite{tropp2006algorithmsL21norm}). 
 Finally,  
$\lambda_1, \lambda_2\geq 0$ are regularization parameters that set the relative weights of the respective regularizers. Assuming that $\phi$ is a convex function, then the optimization (\ref{eq:2unc}) is a convex problem.

The above reasoning has connections with that of robust PCA \cite{robustPCA}. There are, however, considerable differences:  regarding $\bf X$, we explicitly require it to be low-rank and promote self-similarity through suitable patch-based regularization, whereas robust PCA promotes the low-rankness of the image through nuclear norm regularization and does not use spatial regularization. Regarding $\bf S$, we
promote colummn-wise sparsity through $\ell_{2,1}$ regularization,  whereas robust PCA promotes
sparsity of any element of $\bf S$ through  $\ell_{1}$ regularization.

Let ${\bf A} =  [{\bf Z}^T,~{\bf S}^T]^T$
be a $(p+n_b)\times n$ matrix that combines the $(p \times n)$ eigen-images ${\bf Z}$ and the $(n_b \times n)$ outlier matrix $\bf S$. The optimization problem (\ref{eq:2unc}) can be rewritten as
\begin{align}
  \nonumber
  &\widehat{{\bf A}} \in \arg\min_{{\bf A} } 
   \frac{1}{2} \|  {\bf Y} - [{\bf E}, {\bf I}_{n_b}] {\bf A} \|_F^2  + \lambda_1 \phi ([{\bf I}_p, {\bf 0}_{(p \times n_b)}] {\bf A})\\
 & \hspace{1.4cm}+ \lambda_2 \| ([{\bf 0}_{(n_b \times p)}, {\bf I}_{n_b}] {\bf A})^T\|_{2,1},
 \label{eq:A}
\end{align}
where ${\bf I}_a$ denotes an identity matrix of size $a$ and ${\bf 0}_{(a \times b)}$ is the zero matrix of size $a \times b$.  

\subsubsection{Solver}
We solve the  optimization problem \eqref{eq:A} using  CSALSA algorithm \cite{SALSA}, which is an instance of ADMM \cite{ADMM} that was developed to solve convex optimizations with an arbitrary number of terms. CSALSA starts by converting the original  optimization into a constrained one by
using variable splitting:
\begin{align}
\label{eq:2con}
\min_{{\bf A}, {\bf V}_1, {\bf V}_2,{\bf V}_3} & \;\;\; \frac{1}{2} \| {\bf Y} - {\bf V}_1 \|_F^2 + \lambda_1 \phi ({\bf V}_2) + \lambda_2 \| {\bf V}_3^T\|_{2,1} \\\nonumber
{\rm s.t.} &\;\;\; {\bf V}_1= [{\bf E}, {\bf I}_{n_b}] {\bf A}\\\nonumber
           &\;\;\; {\bf V}_2 = [{\bf I}_p, {\bf 0}_{(p \times n_b)}] {\bf A}\\\nonumber
           &\;\;\; {\bf V}_3 =  [{\bf 0}_{(n_b \times p)}, {\bf I}_{n_b}] {\bf A}.
\end{align}
The augmented Lagrangian function of the above  optimization is 
\begin{equation}
\label{eq:2aug}
\begin{array}{ll}
L({\bf A}, &\hspace{-0.4cm} {\bf V}_1, {\bf V}_2,{\bf V}_3, {\bf D}_1,{\bf D}_2,{\bf D}_3)  = \\
  & \frac{1}{2} \| {\bf Y} -  {\bf V}_1 \|_F^2 +  \lambda_1 \phi ({\bf V}_2) + \lambda_2 \| {\bf V}_3^T\|_{2,1} \\
 &+ \frac{\mu_1}{2} \|{\bf V}_1- [{\bf E}, {\bf I}_{n_b}] {\bf A} -{\bf D}_1 \|^2_F \\
& +  \frac{\mu_2}{2} \|{\bf V}_2 - [{\bf I}_p, {\bf 0}_{(p \times n_b)}] {\bf A} -{\bf D}_2 \|^2_F  \\
& + \frac{\mu_3}{2} \|{\bf V}_3-  [{\bf 0}_{(n_b \times p)}, {\bf I}_{n_b}] {\bf A}  -{\bf D}_3  \|^2_F,
\end{array}
\end{equation}
where $\mu_1, \mu_2,\mu_3 > 0$ are the CSALSA penalty parameters. 

The application of CSALSA to  \eqref{eq:2aug} leads to {\bf Algorithm \ref{alg1}}\footnote{\cred{MATLAB codes of our proposed RhyDe method is available in \url{https://github.com/LinaZhuang/RobustHyDenoiser-RhyDe-}} }.
\begin{algorithm}[h]  
    \caption{{\bf R}obust {\bf hy}perspectral {\bf De}noising ({\bf RhyDe})}
    \begin{algorithmic}[1]  
   \label{alg:2}
    \STATE Set $k=0$, choose $\mu_1,\mu_2,\mu_3>0$, ${\bf V}_{1,0}$, ${\bf V}_{2,0}$, ${\bf D}_{1,0}$, ${\bf D}_{2,0}$. 
\REPEAT    
\STATE ${\bf A}_{k+1} = \arg \min_{\bf A} \frac{\mu_1}{2} \| {\bf V}_{1,k}- [{\bf E}, {\bf I}_{n_b}] {\bf A} -{\bf D}_{1,k} \|_F^2 + \frac{\mu_2}{2} \|{\bf V}_{2,k} - [{\bf I}_p, {\bf 0}_{(p \times n_b)}] {\bf A} -{\bf D}_{2,k} \|^2_F 
+ \frac{\mu_3}{2} \|{\bf V}_{3,k}-  [{\bf 0}_{(n_b \times p)}, {\bf I}_{n_b}] {\bf A}  -{\bf D}_{3,k}  \|^2_F$
    \label{alg:A} 
 \vspace{2mm}
\STATE ${\bf V}_{1,k+1}  = \arg \min_{{\bf V}_1}   \frac{1}{2} \| {\bf Y} - {\bf V}_1 \|_F^2 + \frac{\mu_1}{2} \| {\bf V}_1- [{\bf E}, {\bf I}_{n_b}] {\bf A}_{k+1} -{\bf D}_{1,k} \|_F^2$
    \label{alg:V1} 
    \vspace{2mm}  
\STATE  ${\bf V}_{2,k+1}  = \arg \min_{{\bf V}_{2,k+1}}   \lambda_1 \phi ({\bf V}_2) +  \frac{\mu_2}{2} \|{\bf V}_2 - [{\bf I}_p, {\bf 0}_{(p \times n_b)}] {\bf A}_{k+1} -{\bf D}_{2,k} \|^2_F$
    \label{alg:V2} 
    \vspace{2mm}  
    \STATE  ${\bf V}_{3,k+1}  = \arg \min_{{\bf V}_{3,k+1}}   \lambda_2 \| {\bf V}_3^T\|_{2,1}  + \frac{\mu_3}{2} \|{\bf V}_3-  [{\bf 0}_{(n_b \times p)}, {\bf I}_{n_b}] {\bf A}_{k+1}  -{\bf D}_{3,k}  \|^2_F$
    \label{alg:V3} 
    \vspace{2mm}  
\STATE  ${\bf D}_{1,k+1}  =  {\bf D}_{1,k} - ( {\bf V}_{1,k+1}- [{\bf E}, {\bf I}_{n_b}] {\bf A}_{k+1} )$
\vspace{2mm}  
\STATE  ${\bf D}_{2,k+1}  =   {\bf D}_{2,k} - ( {\bf V}_{2,k+1} - [{\bf I}_p, {\bf 0}_{(p \times n_b)}] {\bf A}_{k+1} )$
\vspace{2mm}  
\STATE  ${\bf D}_{3,k+1}  = {\bf D}_{3,k} - ({\bf V}_{3,k+1} -  [{\bf 0}_{(n_b \times p)}, {\bf I}_{n_b}] {\bf A}_{k+1}) $
\vspace{2mm}  
\STATE $k \rightarrow k+1$
\UNTIL{stopping criterion is satisfied.}
    \end{algorithmic}  
    \label{alg1}
    \end{algorithm}  
The optimizations of lines  \ref{alg:A} and  \ref{alg:V1}  are  quadratic problems, whose solutions are given by
\begin{equation}
\begin{array}{ll}
{\bf A}_{k+1} 
= & \big( \mu_1 [{\bf E}, {\bf I}_{n_b}]^H [{\bf E}, {\bf I}_{n_b}] + \mu_2 [{\bf I}_p, {\bf 0}_{p \times n_b}]^H\\
 &[{\bf I}_p, {\bf 0}_{p \times n_b}]
+ \mu_3 [{\bf 0}_{(n_b \times p)}, {\bf I}_{n_b}]^H [{\bf 0}_{(n_b \times p)}, {\bf I}_{n_b}]\big)^{-1}\\
 & \big(  \mu_1 [{\bf E}, {\bf I}_{n_b}]^H ({\bf V}_1 - {\bf D}_1) 
+ \mu_2 [{\bf I}_p, {\bf 0}_{p \times n_b}]^H\\
& ({\bf V}_2 - {\bf D}_2)  
 + \mu_3 [{\bf 0}_{(n_b \times p)}, {\bf I}_{n_b}]^H ({\bf V}_3 - {\bf D}_3) \big),
\end{array}
\end{equation}
(for line 3) and
\begin{equation}
{\bf V}_{1,k+1} = (1+\mu_1)^{-1}  \big[ {\bf Y} + \mu_1 ( [{\bf E}, {\bf I}_{n_b}]{\bf A}_{k+1} + {\bf D}_{1,k}) \big].
\end{equation}
(for line 4).

\noindent
Lines 5 and 6 of Algorithm 1 are optimizations that  are  proximity operators (POs) of $\phi$ applied to ${\bf V}'_{2,k} = [{\bf I}_p, {\bf 0}_{(p \times n_b)}] {\bf A}_{k+1} +{\bf D}_{2,k}$, and of the $\ell_{2,1}$ norm applied to
${\bf V}'_{3,k} = [{\bf 0}_{(n_b \times p)}, {\bf I}_{n_b}] 
{\bf A}_{k+1}+{\bf D}_{3,k}$, respectively. That is
\begin{equation}
\label{eq:2outimg}
{\bf V}_{2,k+1} = \Psi_{\lambda_1 \phi/\mu_2}({\bf V}'_{2,k}), 
\end{equation}  
where 
   \begin{equation}
   \label{eq:subp_Z}
   \Psi_{\lambda \phi}({\bf U}) = \arg \min_{\bf X} \frac{1}{2}\| {\bf X} - {\bf U}\|_F^2 + \lambda \phi({\bf X}),
   \end{equation}
and
\begin{equation}
\label{eq:2v2_obj}
{\bf V}_{3,k+1} = \Psi_{\lambda_2 \|\cdot\|_{2,1}/\mu_3}({\bf V}'_{3,k}), 
\end{equation}  
where, given ${\bf U} = [{\bf u}_1,\dots,{\bf u}_u]$,
   \begin{align}
   \Psi_{\lambda \|\cdot\|_{2,1}}({\bf U})  = & \arg \min_{\bf X} \frac{1}{2}\| {\bf X} - {\bf U}\|_F^2 + \lambda \|{\bf X}^T\|_{2,1}\\
                 = & [\text{soft-vector}({\bf u}_i, \lambda/\mu_3),\, i=1,\dots,n],
                 \label{eq:soft-vector}
   \end{align}
and $\text{soft-vector}({\bf x},\tau) $ is the vector-soft-threshold \cite{combettes2011proximal}
$$
  {\bf x} \mapsto \frac{\max(\|{ \bf x}\|-\tau,{\bf 0})}{\max(\|{ \bf x}\|-\tau,{\bf 0}) + \,\tau}{\bf x}.
$$ 
Actually, ${\bf V}'_{3,k}$ and ${\bf V}'_{2,k}$ can be regarded, respectively,  as an estimate of a noisy outlier image, which only contains information about outliers, and an estimate of noisy eigen-images, respectively, in the $k$-th iteration. Functions $\Psi_{\lambda_1 \phi/\mu_2}({\bf V}'_{2,k})$ and $\Psi_{\lambda_2 \|\cdot\|_{2,1}/\mu_3}({\bf V}'_{3,k})$ play the role of denoisers of the respective  noisy images.

As the subspace spanned by the columns of $\bf E$ is orthogonal,   eigen-images, ${\bf Z}$, lying on the subspace tend to be decorrelated. Taking advantage of the decorrelation of eigen-images, we decouple the function $\phi(\cdot)$ with respective to the eigen-images, that is
\begin{equation}
\phi({\bf Z}) = \sum_{i=1}^p \phi_i({\bf Z}^i),
\end{equation}
where ${\bf Z}^i$ denotes $i$th row of matrix ${\bf Z}$. The solution of \eqref{eq:2outimg} can be decoupled w.r.t. rows of matrix ${\bf V}'_{2,k}$  and can be written as
\begin{equation}
{\bf V}_{2,k+1} = \Psi_{\lambda_1 \phi/\mu_2}({\bf V}'_{2,k}) 
=  \left[
\begin{array}{c}
\Psi_{\lambda_1 \phi_1/\mu_2}\big(({\bf V}'_{2,k})^1 \big) \\
\vdots \\
\Psi_{\lambda_1 \phi_p/\mu_2}\big(({\bf V}'_{2,k})^p \big)
\end{array}
\right]
\end{equation}  
where 
   \begin{equation}
   \label{eq:denoiserPlugged}
   \Psi_{\lambda \phi_i}({\bf u}) = \arg \min_{\bf x} \frac{1}{2}\| {\bf x} - {\bf u}\|_F^2 + \lambda \phi({\bf x})
   \end{equation}
is the so-called denoising operator.

\subsubsection{plug-and-play prior, $\phi(\cdot)$}
\label{sec:ppp}

As \eqref{eq:denoiserPlugged} is a  denoising problem of a single-band image, we can directly apply an available denoiser to the image:
  \begin{equation}
  \label{eq:denoiserPlugged2}
   \Psi_{\lambda \phi_i}({\bf u}) \leftarrow \text{denoiser}({\bf u}),
   \end{equation}
where denoiser$(\cdot)$ is a plugged state-of-the-art denoiser, such as BM3D \cite{BM3D} and WNNM \cite{gu2014weighted}. 
The replacement of \eqref{eq:denoiserPlugged} by \eqref{eq:denoiserPlugged2} is called 
the 
plug-and-play step \cite{Plug-and-Play}.

Here,  we use the plug-and-play  approach  in  solving the subproblem w.r.t. ${{\bf Z}}$.  
Since   those denoisers are not  proximity operators, we do not have a convergence guarantee for the implemented variant of CSALSA.  The convergence of the plug-and-play iterative procedures is currently an active area of research \cite{Plug-and-Play}. In our case, we have systematically observed the convergence of RhyDe and the augmented Lagrangian parameters set to $\mu_i = 1$, for $i=1,2,3$.

After estimating $\widehat{\bf Z}$ and $\widehat{\bf S}$, the denoised image is obtained as follows:
\begin{equation}
\widehat{\bf X} = {\bf E} \widehat{\bf Z} + \widehat{\bf S}.
\end{equation}

\subsection{Denoiser for dealing with Gaussian non-i.i.d. Noise}
\label{sec:noniid}

The spectral covariance of noise, denoted as ${\bf C}_{\lambda} \in \mathbb{R}^{n_b \times n_b}$,  is defined as
\begin{equation}
{\bf C}_{\lambda} = \mathbb{E}[{\bf n}_i {\bf n}_i^T],
\end{equation}
where ${\bf n}_i$ is a generic column of the noise matrix ${\bf N}$ and $\mathbb{E}(\cdot)$ is an expectation operator. 
In the case of the Gaussian i.i.d. noise assumed in Section~\ref{sec:iid}, we have ${\bf C}_{\lambda} = \sigma^2{\bf I}$, where $\sigma^2$ is the variance of the Gaussian distribution.

In real HSIs, however, the  intensity of the noise in each band varies.  For example,
\cite{8937069} illustrates the SNRs for the visible and near-infrared (VNIR) and short-wave infrared (SWIR) channels of the China's GaoFen-5 satellite, as obtained in the laboratory using a solar simulator.
The values of the SNR change as a function of the channel wavelength (see Fig. 9  in \cite{8937069}).   Therefore, real noise in HSIs tends to be non-i.i.d.; that is, it is pixelwise independent but bandwise dependent. 
We assume that ${\bf C}_{\lambda}$  is positive definite and therefore non-singular.
We can convert the non-i.i.d. scenario into an i.i.d. one by using data-whitening:
\begin{equation}
\label{eq:conversion}
\widetilde{\bf Y} := \sqrt{{\bf C}_{\lambda}^{-1}} {\bf Y},
\end{equation}
where $ \sqrt{{\bf C}_{\lambda}}$  and $  {\bf C}_{\lambda}^{-1}$ denote the square root of ${\bf C}_{\lambda}$ and the  inverse of ${\bf C}_{\lambda}$, respectively. 

The noise in the whitened image is standard Gaussian and i.i.d.  since we have 
\begin{equation}
\widetilde{\bf Y} = \sqrt{{\bf C}_{\lambda}^{-1}} {\bf X} + \sqrt{{\bf C}_{\lambda}^{-1}} {\bf N},
\end{equation}
and spectral covariance of noise in the converted image is 
\begin{align}
\nonumber
\widetilde{{\bf C}}_{\lambda} &= \mathbb{E}[ \sqrt{{\bf C}_{\lambda}^{-1}}{\bf n}_i  (\sqrt{{\bf C}_{\lambda}^{-1}} {\bf n}_i)^T] 
= \sqrt{{\bf C}_{\lambda}^{-1}} \mathbb{E}[ {\bf n}_i  {\bf n}_i^T ] \sqrt{{\bf C}_{\lambda}^{-1}}^T \\
&= \sqrt{{\bf C}_{\lambda}^{-1}} {\bf C}_{\lambda} \sqrt{{\bf C}_{\lambda}^{-1}}^T = {\bf I}.
\end{align}
As a pre-processing step,   \eqref{eq:conversion} converts the noise to 
be standard Gaussian and i.i.d..
  Therefore, the converted image $\widetilde{\bf Y}$ can be denoised as discussed in Section~\ref{sec:iid}. Finally, a clean image is recovered as
\begin{equation}
\widehat{\bf X} = \sqrt{{\bf C}_{\lambda}} \widehat{ \widetilde{\bf X}},
\end{equation}
where $\widehat{ \widetilde{\bf X}}$ is the estimated clean version of image $\widetilde{\bf Y} $.

\subsection{Anomaly Detection}

We propose an anomaly detector derived from the estimate of the outlier matrix ${\widehat{{\bf S}}}$ in (\ref{eq:2unc}). The proposed detector can be expressed as
\begin{equation}
\label{eq:detector}
r_i= \|{\hat {\bf s}}_i\|_2,~~i=1, \dots, n,
\end{equation}
where 
${\hat {\bf s}}_i$ is the $i$-th column of outlier matrix ${\widehat{{\bf S}}}$. If $r_i$ is larger than a given threshold, then the $i$-th pixel is classified  as an anomalous pixel.

A low-rank representation of the background and the structured sparsity of the anomaly are also exploited in LRASR \cite{ZWu}, which is distinct from our RhyDe in terms of the composition of the dictionary, the implementation of the low-rank regularization, and the noise reduction. RhyDe assumes that the spectral vectors of the background live in an orthogonal subspace, implements explicit low-rank representation, and promotes self-similarity of the representation coefficients in order to remove noise. 
LRASR adopts a background dictionary composed of selected image pixels and enforces the low rank and sparsity by applying, respectively, the nuclear norm and $\ell_1$ norm regularizations  to the  representation coefficients without taking any further steps to remove the noise.
 
\section{Experiments with semi-real images}
\label{sec:simexp}
 \subsection{Simulation of Noisy Images and Comparisons}
\subsubsection*{(a) Simulation of noisy images}

 \begin{figure}[htbp]
\centering
\includegraphics[width=8cm]{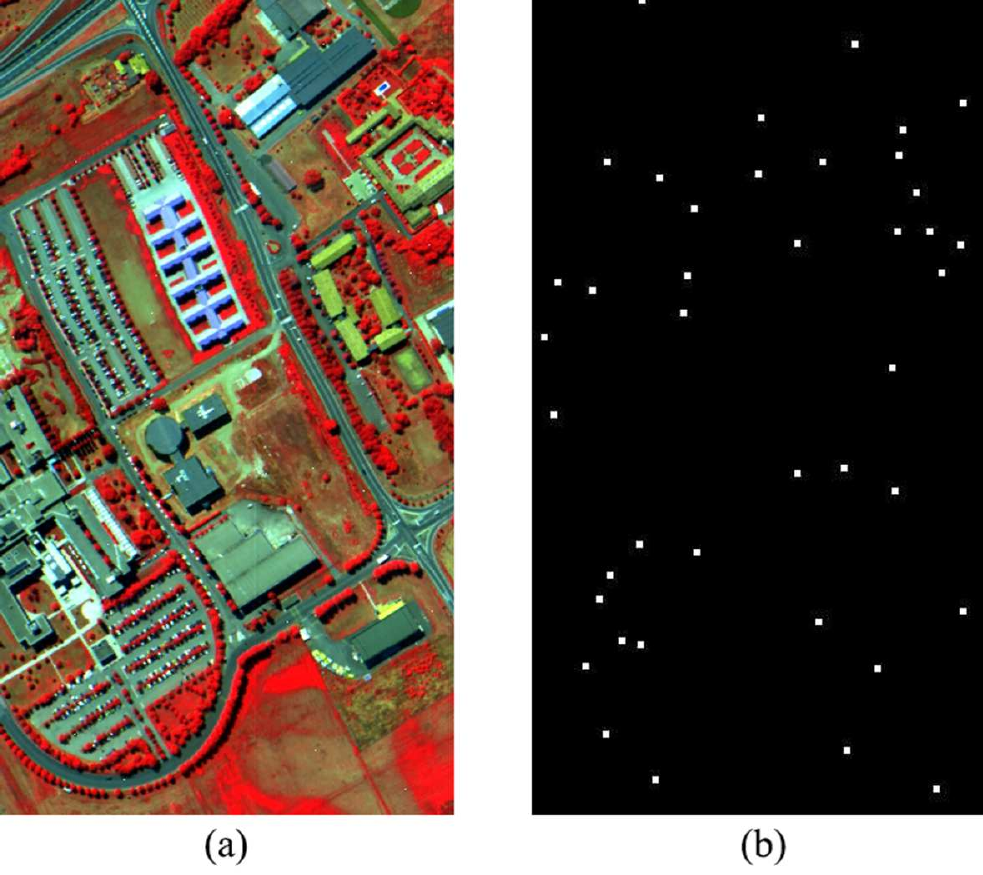}
\caption{(a) Clean Pavia University scene, and (b) Groundtruth of outliers (0.02\%). }
\label{fig:pavia_clean_noisy_est2}
\end{figure}

 \begin{figure}[htbp]
\centering
\includegraphics[width=8cm]{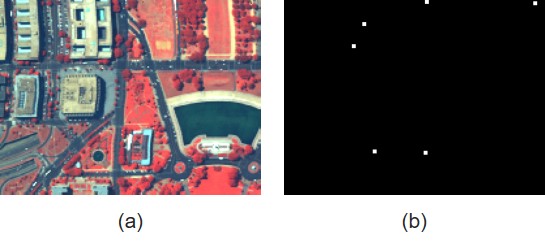}
\caption{(a) Clean Washington DC Mall scene, and (b) Groundtruth of outliers (0.02\%). }
\label{fig:DC_clean}
\end{figure}

Two semi-real hyperspectral datasets  were simulated by adding Gaussian noise and anomalous pixels to the Pavia University image\footnote{  Available in \url{http://www.ehu.eus/ccwintco/index.php?title=Hyperspectral_Remote_Sensing_Scenes}.} and Washington DC Mall image\footnote{  Available in \url{  https://engineering.purdue.edu/~biehl/MultiSpec/hyperspectral.html}} as follows.
\begin{itemize}
\item 
To simulate clean images, 28 very-low SNR bands of the Pavia University data were removed and the 70 bands with relatively high SNRs from the Washington DC Mall data were used.
\item 
The remaining spectral vectors were then projected onto the signal subspace (with  dimension 5) that was learned by singular value decomposition (SVD). Most of the noise was removed by this projection. Fig. \ref{fig:pavia_clean_noisy_est2}-(a) and Fig. \ref{fig:DC_clean}-(a) show  false-color
images produced using three bands taken from the clean image.
\item
In order to simulate outliers, we randomly selected 0.02\% of the pixels and then 
replaced them (Fig.\ref{fig:pavia_clean_noisy_est2}-(b) and Fig. \ref{fig:DC_clean}-(b)) with the spectral signatures
of the minerals sillimanite HS186.3B and ammonio-jarosite (taken from the USGS spectral library\footnote{Available in    \url{https://pubs.er.usgs.gov/publication/ds1035}}) in the case of the Pavia University image and the Washington DC Mall image,   respectively. By doing this, two clean images containing outliers were generated. 
\item
Finally, Gaussian-independent noise together with band-dependent variance was added to simulate noisy images in the form ${\bf n}_i \sim \mathcal{N}({\bf 0}, {\bf D}^2)$ where ${\bf{D}} \in \mathbb{R}^{n_b \times n_b}$ is a diagonal matrix with diagonal elements sampled from a uniform distribution $U(0,u)$ and $u \in \{0.12, 0.09, 0.065, 0.04\}$ for cases 1-4, respectively, in Table \ref{tab:msnr}. 
\end{itemize}

 \subsubsection*{(b) Comparisons}
We made comparisons with other state-of-the-art method of multiband and hyperspectral denoising, namely NAILRMA \cite{he2015hyperspectral}, LRTV \cite{LRTV},  FastHyDe\footnote{\cred{Matlab codes of FastHyDe method is available in \url{https://github.com/LinaZhuang/FastHyDe_FastHyIn}}} \cite{FastHyDe}, and GLF\footnote{\cred{Matlab codes of GLF method is available in \url{https://github.com/LinaZhuang/HSI-denoiser-GLF}}} \cite{GLF}. NAILRMA is a patch-wise low-rank matrix approximation designed for addressing independent Gaussian noise with band-dependent variance. LRTV is a pixel-wise total variation- (TV)-regularized low-rank matrix-factorization method that models observations as the sum of a spectrally low-rank and spatially smooth clean image, sparse noise (using the  $\ell_1$ norm), and Gaussian noise. FastHyDe assumes clean spectral vectors that live in a low-dimensional subspace, and corresponding subspace coefficient components that are self-similar and uncorrelated: these can be denoised component-by-component by non-local patch-based single-band denoisers. Similar to FastHyDe, GLF represents HSI in a low-dimensional subspace, but denoises the subspace coefficients by tensor factorization.

Since the simulated noise is band-variant and FastHyDe, GLF, and RhyDe assume i.i.d noise, the observed data are whitened (as discussed in Section~\ref{sec:noniid}) before these denoisers  are applied.

To make a quantitative assessment, the peak signal-to-noise (PSNR) index and the structural similarity (SSIM) index \cite{FastHyDe} of each band are calculated.  The corresponding mean PSNR (MPSNR) and mean SSIM (MSSIM) are given in Table \ref{tab:msnr} for comparison. 
As MPSNR and MSSIM do not reflect the reconstruction errors in the third (spectral) dimension, we added two image quality metrics:
3D-PSNR and mean spectral angle mapper (MSAM). Their definitions are given as follows:
\begin{equation}
\text{3D-PSNR} = 10 \log_{10}\frac{x_{max}^2}{ \| {\bf X} - {\widehat {\bf X}}\|_F^2},
\end{equation}
where $x_{max}$ is the maximum possible entry value of  matrix ${\bf X}$.
\begin{equation}
\text{MSAM} = \frac{1}{n}\sum_{i=1}^n \cos^{-1} \frac{{\bf x}_i^T {\hat {\bf x}}_i}{\| {\bf x}_i\| \| {\hat {\bf x}}_i \|},
\end{equation}
where ${\bf x}_i$ and ${\hat{\bf x}}_i$ are a clean spectral vector of $i$th pixel and its estimated vector, respectively.

To compare anomaly detectors for comparison, we used two groups: classical detectors (global RX \cite{RX},  local RX \cite{RX}, OSP global RX \cite{OSP-based}, and  OSP local RX \cite{OSP-based}) and a group of recently  developed ones (CRD \cite{NRS_WeiLi},  BSJSBD \cite{BSJSBD},  and LRASR) \cite{ZWu}) that are based on structured sparsity and are either state-of-the-art or competitive with the state-of-art methods.

\subsection{Denoising Performance}
\label{sec:parametersetting}

\begin{table*}[htbp]
  \centering
 \caption{Quantitative assessment of different denoising algorithms applied to semi-real datasets.}
    \begin{tabular}{cccccccc}
    \toprule
          &       & Noisy Image & NAILRMA & LRTV  & FastHyDe & GLF   & RhyDe \\
       &  &  &  \cite{he2015hyperspectral}  &  \cite{LRTV}   &  \cite{FastHyDe}  & \cite{GLF}  &  (proposed)  \\
  \midrule
    \multicolumn{8}{c}{Pavia University datasets} \\
   \midrule
    \multirow{5}[2]{*}{case 1} & MPSNR (dB) & 26.31 & 41.54 & 40.13 & 49.15 & 49.78 & 49.25 \\
          & 3D PSNR  (dB) & 24.49 & 43.74 & 35.35 & 50.77 & 51.42 & 50.94 \\
          & MSSIM & 0.5214 & 0.9824 & 0.9624 & 0.9960 & 0.9965 & 0.9961 \\
          & MSAM  & 25.91 & 2.41  & 7.50  & 1.04  & 0.94  & 1.03 \\
          & Time (s) & -     & 290   & 622   & 14    & 129   & 129 \\
    \midrule
    \multirow{5}[2]{*}{case 2} & MPSNR (dB) & 28.08 & 42.40 & 40.88 & 48.36 & 49.27 & 48.52 \\
          & 3D PSNR  (dB) & 27.10 & 44.63 & 37.67 & 49.87 & 50.78 & 50.08 \\
          & MSSIM & 0.6045 & 0.9853 & 0.9679 & 0.9952 & 0.9961 & 0.9954 \\
          & MSAM  & 20.16 & 2.18  & 5.77  & 1.14  & 1.01  & 1.13 \\
          & Time (s) & -     & 291   & 627   & 14    & 127   & 130 \\
    \midrule
    \multirow{5}[2]{*}{case 3} & MPSNR (dB) & 30.71 & 43.93 & 42.08 & 50.09 & 50.68 & 50.26 \\
          & 3D PSNR  (dB) & 29.83 & 46.18 & 40.03 & 51.98 & 52.52 & 52.24 \\
          & MSSIM & 0.7103 & 0.9901 & 0.9745 & 0.9968 & 0.9972 & 0.9969 \\
          & MSAM  & 15.28 & 1.82  & 4.41  & 0.93  & 0.84  & 0.92 \\
          & Time (s) & -     & 291   & 616   & 14    & 123   & 130 \\
    \midrule
    \multirow{5}[2]{*}{case 4} & MPSNR (dB) & 34.47 & 48.74 & 44.37 & 54.07 & 54.53 & 54.44 \\
          & 3D PSNR  (dB) & 33.20 & 50.88 & 44.46 & 55.87 & 56.31 & 56.49 \\
          & MSSIM & 0.8123 & 0.9956 & 0.9876 & 0.9986 & 0.9988 & 0.9987 \\
          & MSAM  & 10.72 & 1.17  & 2.46  & 0.59  & 0.55  & 0.59 \\
          & Time (s) & -     & 283   & 634   & 14    & 130   & 127 \\
    \midrule
    \multicolumn{8}{c}{Washington DC Mall datasets} \\
    \midrule
    \multirow{5}[2]{*}{case 1} & MPSNR (dB) & 32.53 & 45.40 & 40.43 & 52.32 & 53.11 & 52.32 \\
          & 3D PSNR  (dB) & 30.02 & 46.97 & 38.26 & 52.85 & 53.82 & 52.87 \\
          & MSSIM & 0.7637 & 0.9889 & 0.9743 & 0.9963 & 0.9975 & 0.9963 \\
          & MSAM  & 9.40  & 1.08  & 3.20  & 0.53  & 0.47  & 0.53 \\
          & Time (s) & -     & 27    & 24    & 2     & 18    & 17 \\
    \midrule
    \multirow{5}[2]{*}{case 2} & MPSNR (dB) & 32.90 & 46.43 & 42.98 & 51.40 & 52.27 & 51.42 \\
          & 3D PSNR  (dB) & 31.57 & 47.64 & 42.44 & 51.93 & 52.55 & 51.94 \\
          & MSSIM & 0.7930 & 0.9904 & 0.9825 & 0.9967 & 0.9976 & 0.9967 \\
          & MSAM  & 7.91  & 1.01  & 1.97  & 0.59  & 0.54  & 0.59 \\
          & Time (s) & -     & 24    & 25    & 2     & 17    & 17 \\
    \midrule
    \multirow{5}[2]{*}{case 3} & MPSNR (dB) & 36.17 & 49.67 & 47.43 & 54.89 & 55.54 & 54.97 \\
          & 3D PSNR  (dB) & 34.64 & 51.47 & 49.15 & 56.75 & 57.42 & 56.81 \\
          & MSSIM & 0.8573 & 0.9950 & 0.9923 & 0.9987 & 0.9990 & 0.9987 \\
          & MSAM  & 5.57  & 0.66  & 0.83  & 0.35  & 0.32  & 0.35 \\
          & Time (s) & -     & 28    & 25    & 2     & 18    & 17 \\
    \midrule
    \multirow{5}[2]{*}{case 4} & MPSNR (dB) & 39.86 & 52.04 & 50.74 & 56.44 & 57.15 & 56.53 \\
          & 3D PSNR  (dB) & 38.69 & 53.07 & 51.76 & 56.27 & 56.97 & 56.33 \\
          & MSSIM & 0.9268 & 0.9968 & 0.9961 & 0.9988 & 0.9991 & 0.9988 \\
          & MSAM  & 3.51  & 0.55  & 0.61  & 0.37  & 0.33  & 0.36 \\
          & Time (s) & -     & 29    & 26    & 2     & 18    & 17 \\
    \bottomrule
    \end{tabular}%
  \label{tab:msnr}%
\end{table*}%

\begin{figure*}[htbp]
\centering
\includegraphics[width=18cm]{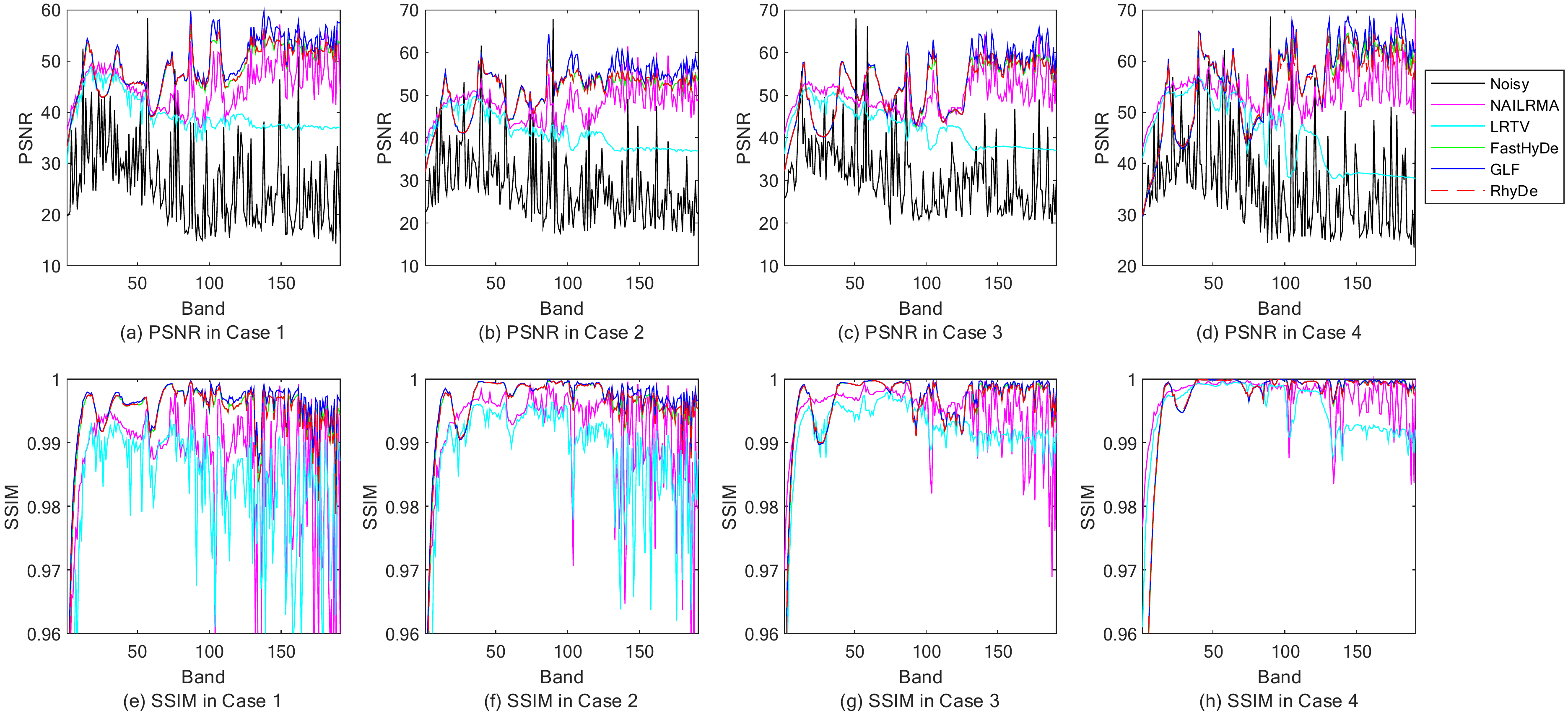}
\caption{PSNR (a-d) and SSIM (e-h) of each band of denoised Pavia University images in Cases 1-4. }
\label{fig:psnr_ssim_curve_pavia}
\end{figure*}

\begin{figure*}[htbp]
\centering
\includegraphics[width=18cm]{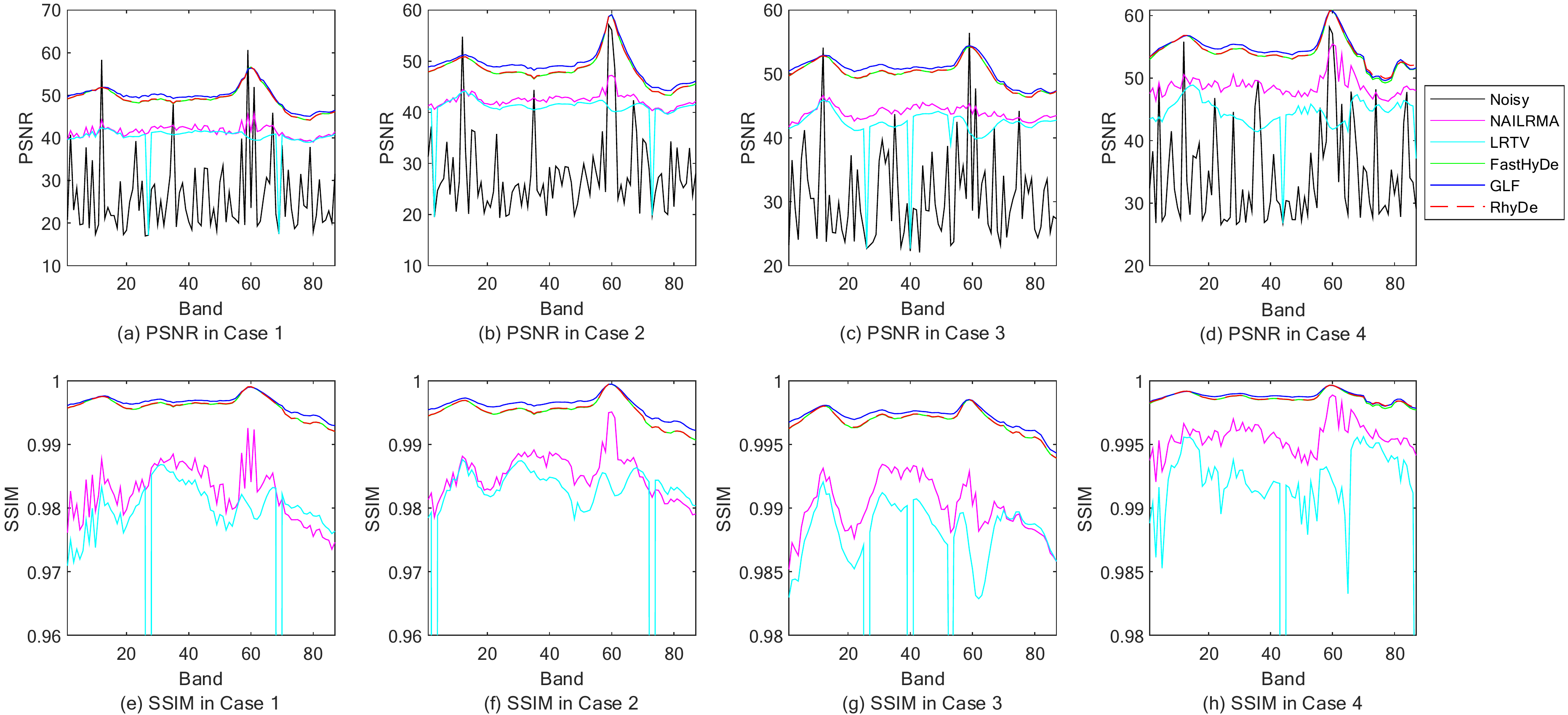}
\caption{PSNR (a-d) and SSIM (e-h) of each band of denoised Washington DC Mall   images in Cases 1-4. }
\label{fig:psnr_ssim_curve_DC}
\end{figure*}

 \begin{figure*}[htbp]
\centering
\includegraphics[width=17cm]{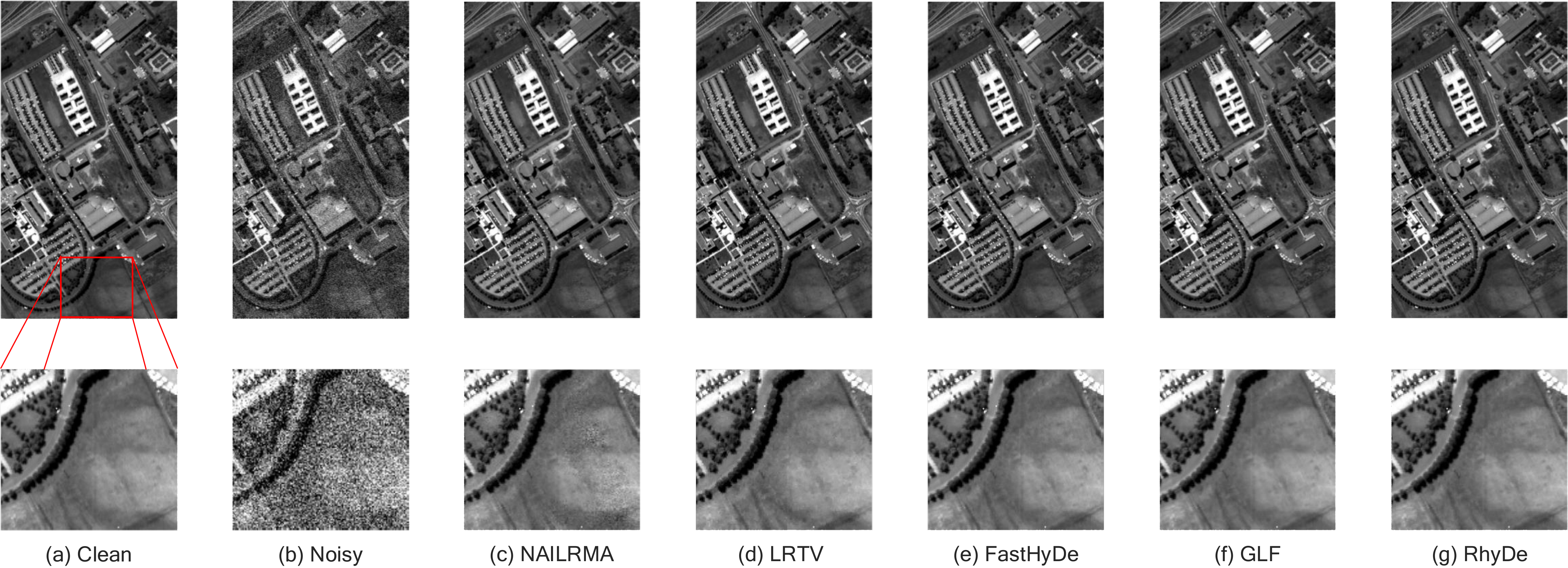}
\caption{ Clean and denoised band 24 of simulated Pavia University scene  (MPSNR = 30.71 dB) in Case 3.}
\label{fig:sim_denoisedImgs}
\end{figure*}

 \begin{figure*}[htbp]
\centering
\includegraphics[width=17cm]{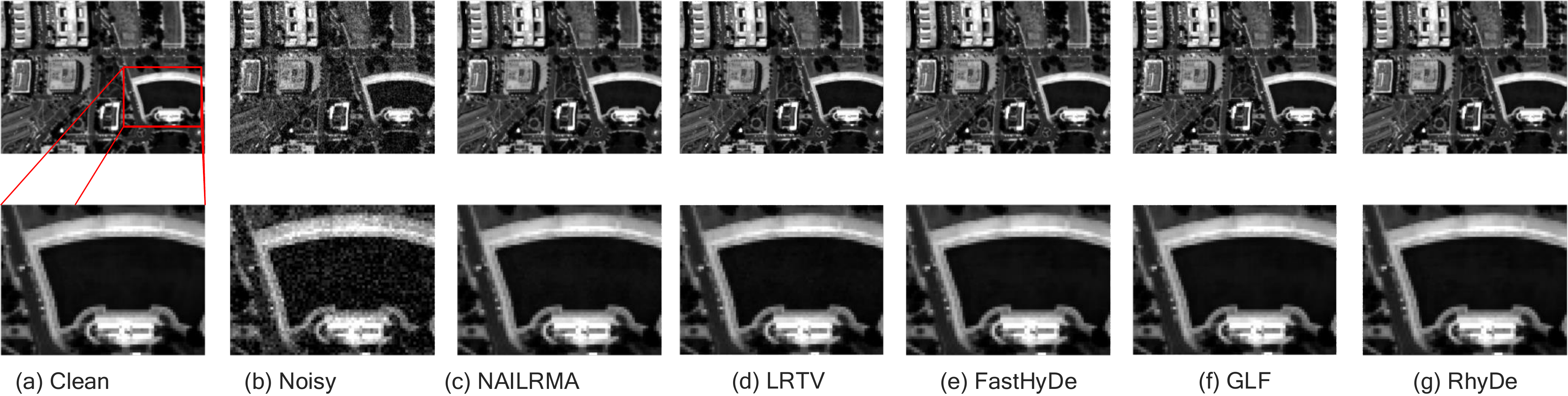}
\caption{ Clean and denoised band 26 of simulated Washington DC Mall scene  (MPSNR = 36.17 dB) in Case 3.}
\label{fig:sim_denoisedImgs_zoom_DC}
\end{figure*}

  \begin{figure*}[htbp]
\centering
\includegraphics[width=17.5cm]{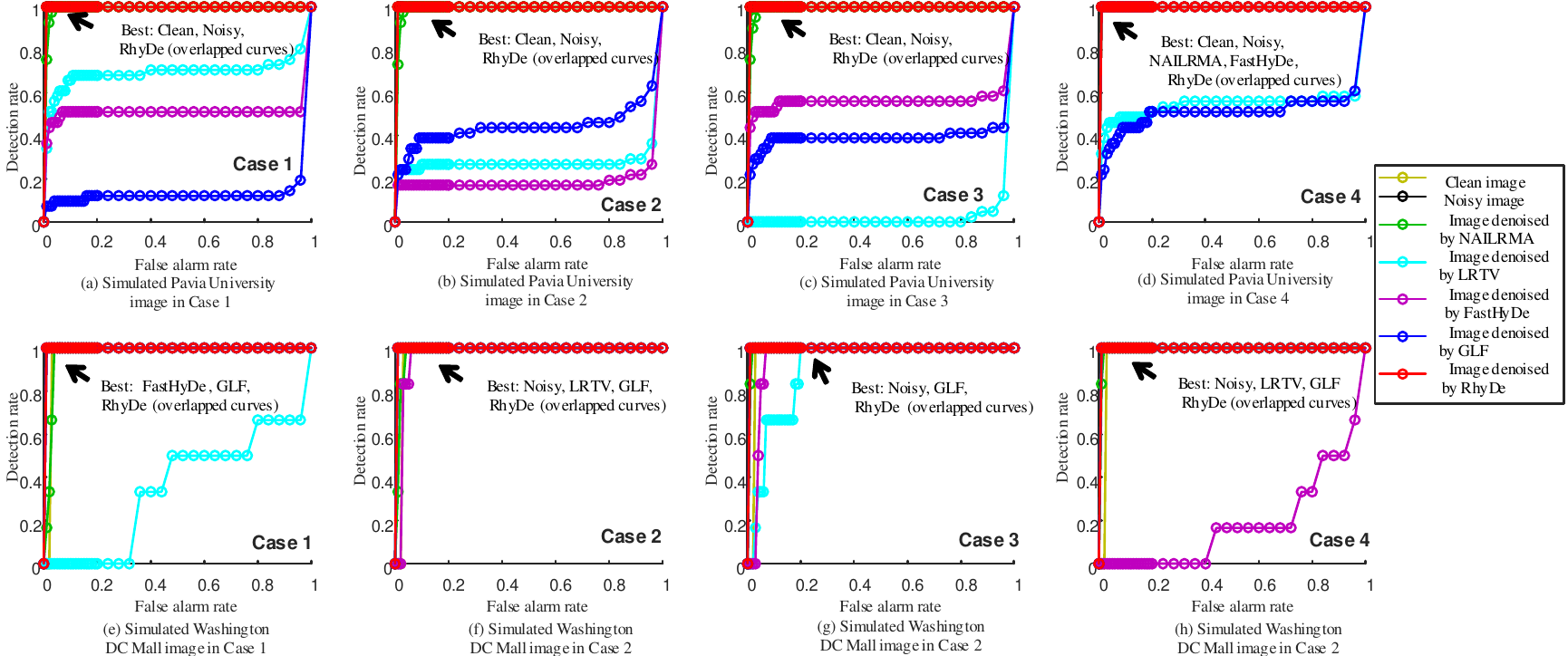}
\caption{ROCs of Global RX detector applying to original and denoised images of simulated Pavia University images in Cases 1-4 (a-d) and simulated Washington DC Mall images in Cases 1-4 (e-h).}
\label{fig:pavia_DC_GRXafterDifDenoiser}
\end{figure*}

\begin{figure*}[htbp]
\centering
\includegraphics[width=18cm]{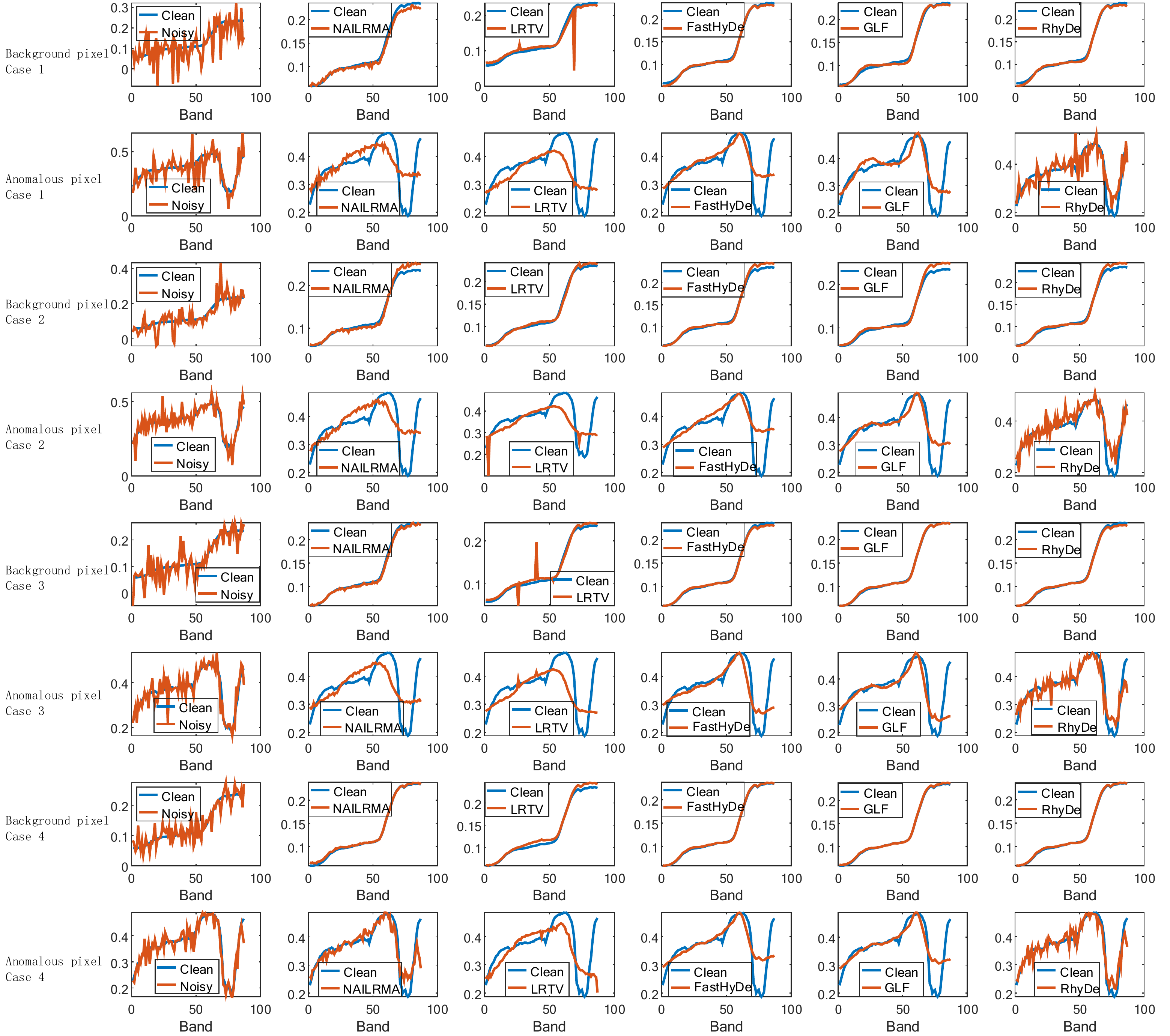}
\caption{Denoised spectral signatures of   background pixels (in the odd-rows)
and
   anomalous pixels (in the even-rows)
in simulated Pavia University scene   in Cases 1-4. Note that the noise in the anomalous pixel is not removed completely by RhyDe since our main objective w.r.t. anomalies is to keep them rather than to denoise them and our output result is $\widehat{{\bf Z}}+\widehat{{\bf S}}$. }
\label{fig:pavia_spectrum_normal_outlier_allcases}
\end{figure*}

We now discuss the parameter settings for the various algorithms. The subspace dimension used as the input to LRTV, FastHyDe, GLF, and RhyDe was set to 5, which was the true value. Since the noise in simulated images is band-dependent, we applied data whitening before denoising; thus, the parameter noise covariance matrix, ${\bf C}_{\lambda}$ was set to the identity matrix ${\bf I}$ for FastHyDe, GLF, and RhyDe.
The RhyDe parameters, $\lambda_1$ and $\lambda_2$, are assigned adaptive to estimate of noise, ${\bf C}_{\lambda}$  (see Section~\ref{sec:parameter} for further details).
The parameters for NAILRMA and LRTV were hand-tuned to ensure optimal performance.

The denoising performance of different algorithms in terms of MPSNR, 3D-PSNR, MSSIM, MSAM, and computation time (in seconds)
is reported in Table \ref{tab:msnr}. The results indicate that the strategy used in FastHyDe, GLF, and RhyDe, i.e., exploiting self-similarity among non-local patches, works better than that used in the patch-wise NAILRMA and pixel-wise LRTV. 
Fig. \ref{fig:psnr_ssim_curve_pavia} and Fig. \ref{fig:psnr_ssim_curve_DC} show PSNR and SSIM values for each band of the denoised Pavia University images and Washington DC Mall images, respectively.
The quality of the reconstructed bands  may also be inferred  from Fig. \ref{fig:sim_denoisedImgs} and  Fig. \ref{fig:sim_denoisedImgs_zoom_DC}. 
It can be seen that all of the denoising methods are capable of reducing the noise in the imagery. 

We used the denoised images for subsequent anomaly detection.  
Fig. \ref{fig:pavia_DC_GRXafterDifDenoiser} shows the effect of the denoising on this detection. The performance of an outlier detector is usually evaluated by using a receiver operating characteristic (ROC) curve, which plots the detection rate against the false alarm rate. Fig. \ref{fig:pavia_DC_GRXafterDifDenoiser} illustrates the ROC curves for the application of Global RX to clean images, noisy images, and denoised images. Global RX was chosen for this evaluation since it is  a representative outlier detector that requires no input parameters and provides a fair environment for making comparisons. It can be seen that, in some of the denoised images, anomalous targets were detected with higher false alarm rates than in the original image even though the denoised images have much higher MPSNR and MSSIM values than the original noisy images. This happens because MPSNR and MSSIM are global image-quality criteria and may not be able to express the degradation of a very small number of pixels. The very high false alarm rates for some denoised images (Fig. \ref{fig:pavia_DC_GRXafterDifDenoiser})  imply that information about rare pixels is corrupted after denoising. 
  This is confirmed by the results shown in Fig. \ref{fig:pavia_spectrum_normal_outlier_allcases}, where the denoised spectral signatures of anomalous pixels and background pixels can be seen. It can be seen that the spectral vectors of background pixels (the odd-numbered rows) were recovered well by all the denoising methods, whereas the spectral vectors of the anomalous pixels (the even-numbered rows) were degraded by all of the methods except RhyDe.

In presence of rare pixels are present, the denoising step may compromise the futuresubsequent detection of those pixels. The reason for this is that the low-rank representation of the image (used in NAILRMA, LRTV, FastHyDe, and GLF) will lead toleads to the a risk of losing the information ofassociated with rare pixels. In contrast with this scenario, RhyDe is able to preserve anomalous pixels (see the even-numbered rows in Fig. \ref{fig:pavia_spectrum_normal_outlier_allcases}) since it represents rare pixels as a column-wise sparse matrix. Note that both  LRTV and RhyDe include an outlier matrix, ${\bf S}$, in their observation models. LRTV assumes that the structure of ${\bf S}$ is randomly sparse and promoted by $\ell_1$ norm, whereas RhyDe assumes that ${\bf S}$ is
column-wise sparse and promoted by the $\ell_{2,1}$ norm. Given that the outlier matrix represents pixels of rare materials, the assumption of a column-wise structure used in RhyDe is more appropriate in this scenario.

\subsection{Parameter Analysis of RhyDe}
 \label{sec:parameter}
 
  \begin{figure*}[htbp]
\centering
\includegraphics[width=18cm]{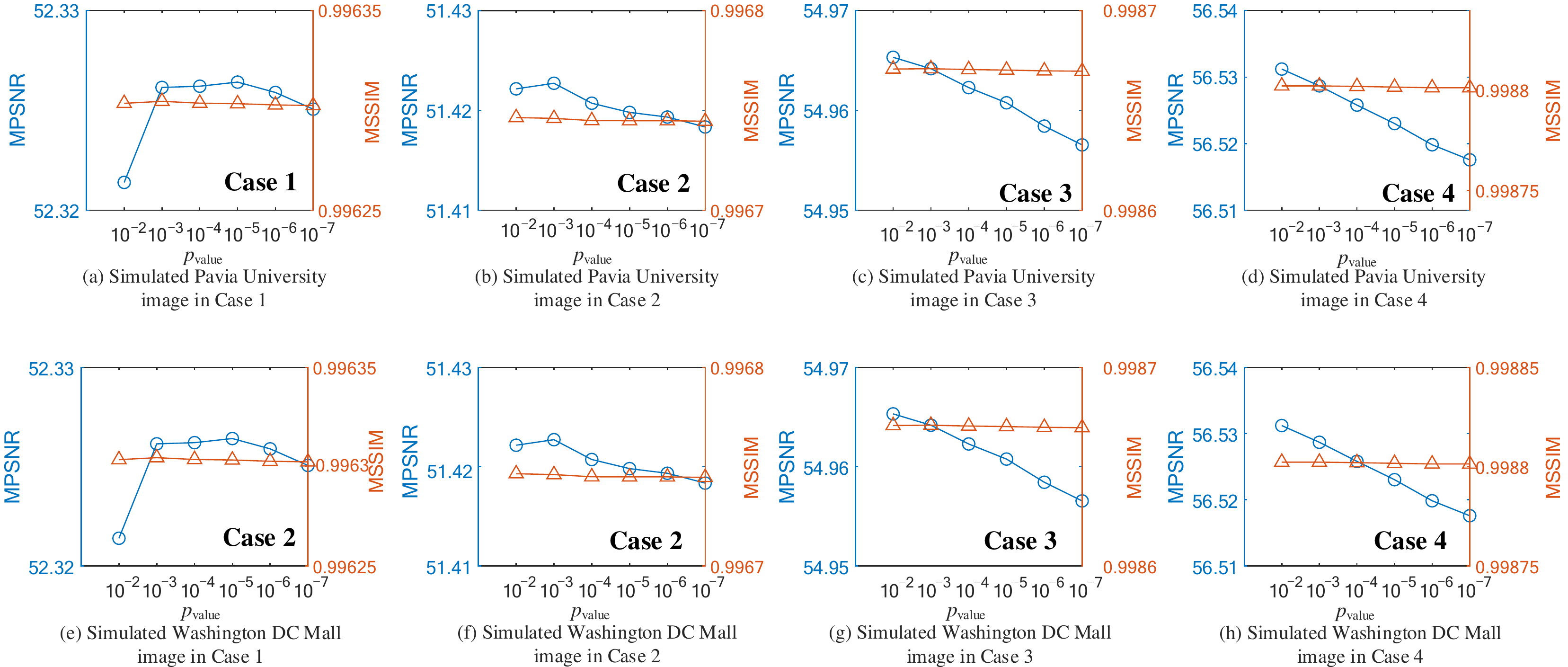}
\caption{The impact of the regularization parameter, $ \lambda_2  =  \sqrt{\text{chi2inv}(1-p_{\text{value}},n_b) }$,  on the denoising performance of Pavia University images (a-d) and Washington DC Mall images (e-h) in terms of MPSNR (in dB)  and MSSIM. }
\label{fig:ParameterAnalysisRhyde}
\end{figure*}

The RhyDe method was proposed for denoising HSIs by solving the optimization problem in \eqref{eq:2unc}, which includes the use of two parameters, $\lambda_1$ and $\lambda_2$. The $\lambda_1$ is the parameter of regularization $\phi({\bf Z})$ tailored to self-similar eigen-images. More specifically, the value of $\lambda_1$ is related to the variances of the Gaussian noise in the eigen-images ${\bf Z}$, which occur when  plug-and-play denoisers \cite{Plug-and-Play}  are used to solve the subproblem \eqref{eq:subp_Z}. Since the variance of the Gaussian noise in each band of the original image can be estimated (for example, by HySime \cite{Hysime}) and the subspace projection is a linear transform, the variance of the Gaussian noise in the projected eigen-images can be estimated easily. 
For example, given the spectral covariance of the noise in the original image, ${\bf C}_{\lambda}$ (estimated by HySime), the    noise variance in the $i$th eigen-image is obtained by computing ${\bf e}_i^T  {\bf C}_{\lambda} {\bf e}_i$ (where ${\bf e}_i$ is the $i$th eigenvector).
Therefore, we can automatically set the value of parameter $\lambda_1$, which is adaptive to the estimate of the noise.

Parameter $\lambda_2$ controls the column-wise sparsity of matrix ${\bf S}$. 
If the HSI is noise-free, then the matrix ${\bf S}$ is supposed to be a column-wise sparse matrix, where only a few non-sparse columns contain  outlier components.
If the HSI is corrupted by Gaussian noise, then the matrix ${\bf S}$, which includes both outlier components and the Gaussian noise (distributed densely in ${\bf S}$), is non-sparse.
In order to exclude Gaussian noise and include only outlier components in matrix ${\bf S}$, the regularization $\| {\bf S} \|_{2,1}$ is added to the objective function. The regularization parameter, $\lambda_2$, should be adaptive to column-wise Gaussian noise intensity.
%
%
In all of the experiments, we whitened the images and then simply set  
$ \lambda_2  =  \sqrt{\text{chi2inv}(1-p_{\text{value}},n_b) }$, where $\text{chi2inv}(\cdot )$   computes the inverse of the chi-square cumulative distribution function with   degrees of freedom specified by $n_b$ for the corresponding probability, $1-p_{\text{value}}$. 
  The reason for this setting was that the distribution of the sum of the squares of $n_b$ independent standard Gaussian noise  is a chi-square  distribution.
 For all experiments carried out in this study, we empirically set $p_{\text{value}} = 10^{-2}$, which meant that the probability of observing a test statistic at least as extreme in this chi-square  distribution was $1- p_{\text{value}} =99\%$.
 Fig. \ref{fig:ParameterAnalysisRhyde} shows the impact of $p_{\text{value}}$ on the denoising performance for all of the simulated images in terms of MPSNR  and MSSIM. Note that the values marked in y-axis indicate  that the proposed RhyDe is robust to the value of   $p_{\text{value}}$.
 
\subsection{Numerical Convergence of the RhyDe}

  \begin{figure}[htbp]
\centering
\includegraphics[width=8cm]{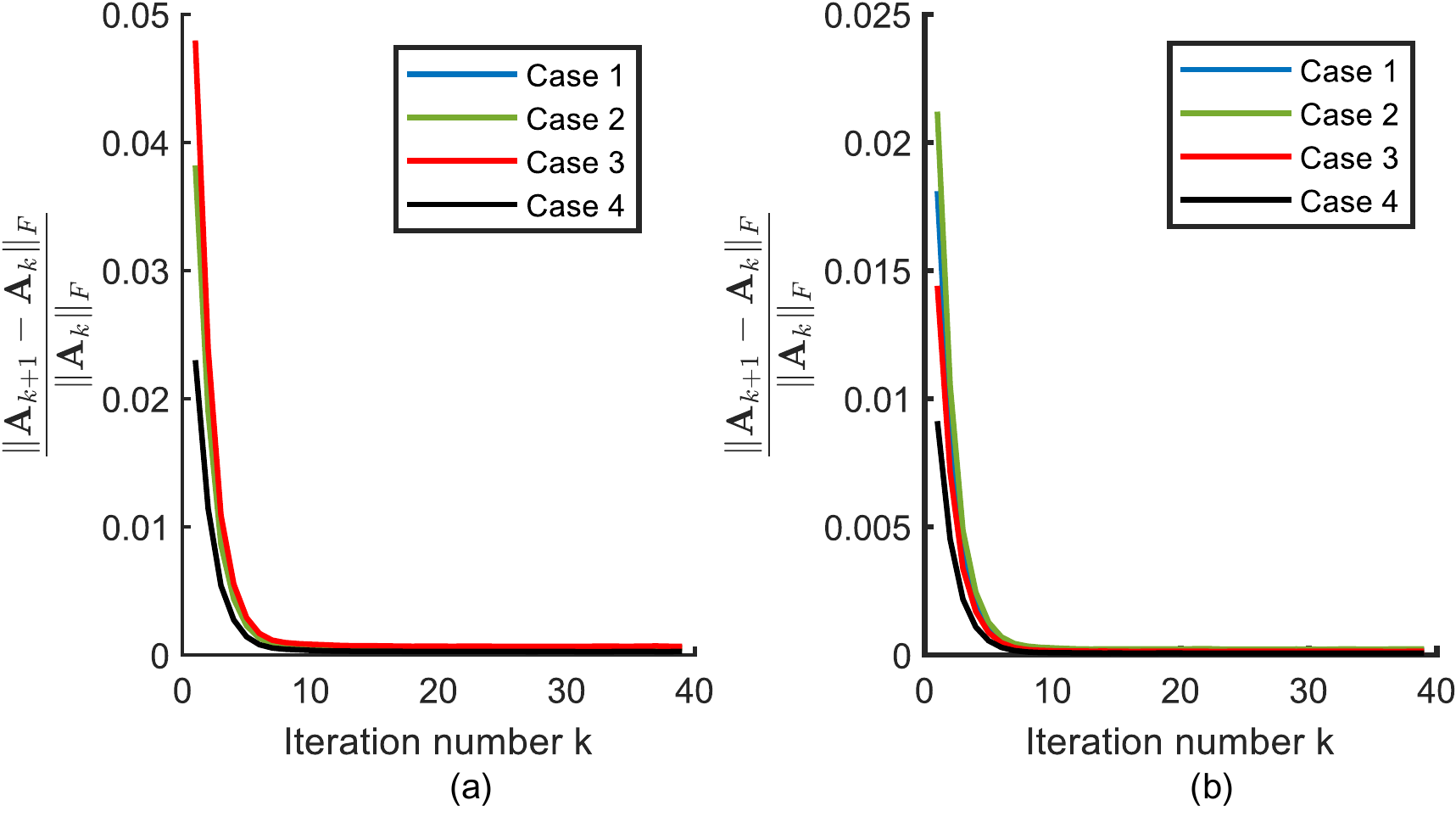}
\caption{Relative change of ${\bf A} = [{\bf Z}^T,~{\bf S}^T]^T$ versus the iteration number of RhyDe denoising Pavia University data in (a) and Washington DC Mall data in (b). }
\label{fig:ConvergenceOfRhyDe}
\end{figure}

The plugged denoiser, BM3D, does not include an explicit convex regularizer; thus, the proposed RhyDe is a  nonconvex optimization problem, and it is hard to prove its convergence. 
The BM3D was chosen as the plugged denoiser in our denoising framework as it is a state-of-the-art single-band denoiser that is user-friendly (requiring only one parameter, the variance of Gaussian noise), and very fast. There is still  room for improving the proposed RhyDe (in terms of the image quality) by using   better-plugged single-band denoisers. However, taking into account  the trade-off between the denoising accuracy and speed, we used BM3D to speed up the iterative framework of RhyDe.
An empirical analysis for the convergence of RhyDe plugged with BM3D is given in Fig. \ref{fig:ConvergenceOfRhyDe}, in which the curves showing the relative changes in ${\bf A}$ during iteration converge after eight iterations. The convergence of RhyDe plugged with BM3D can be numerically guaranteed.

\subsection{Anomaly Detection Performance}
 
\begin{figure}[htbp]
\centering
\includegraphics[width=8.5cm]{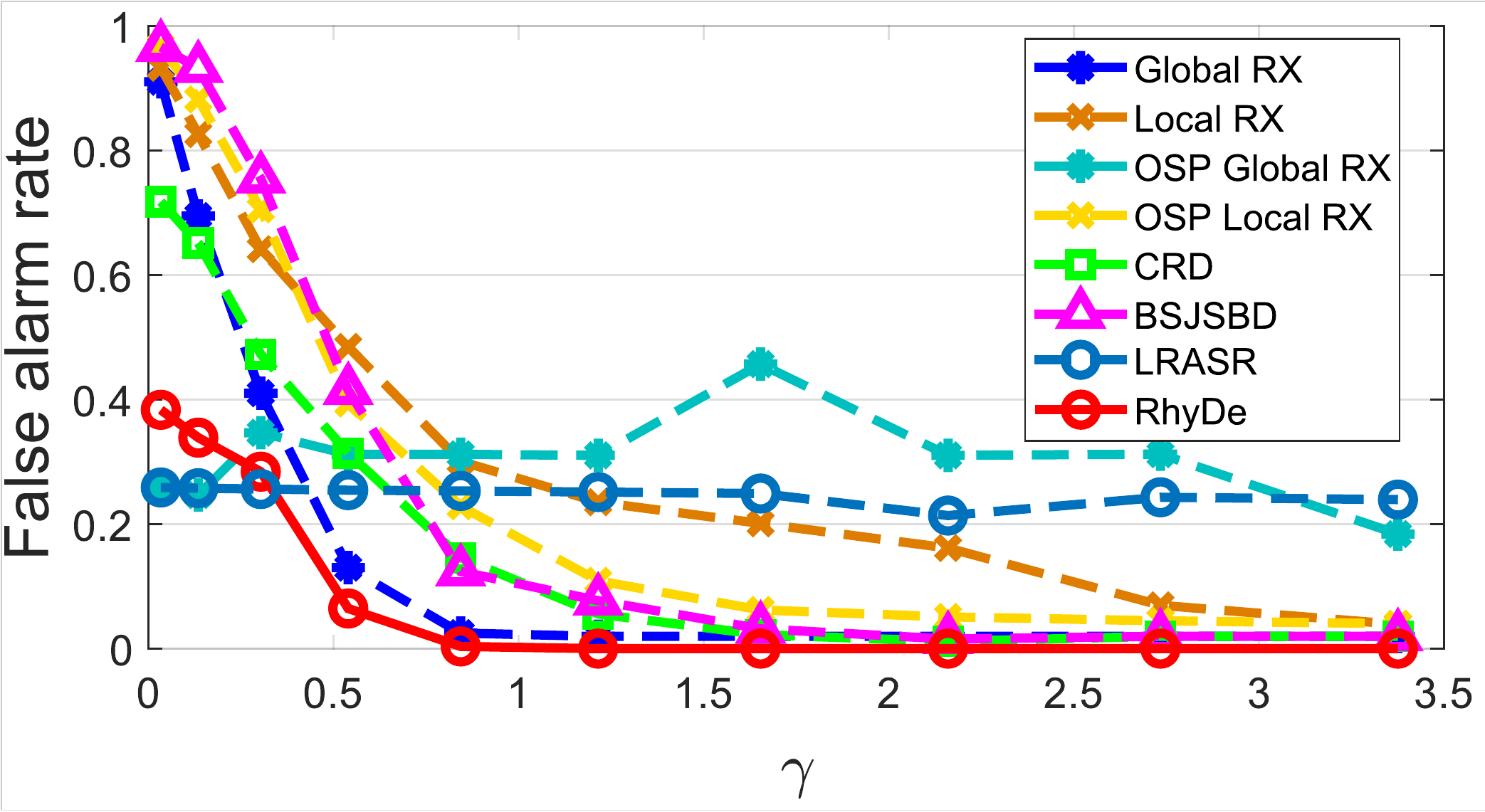}
\caption{ False alarm rate as a function of the relative power of the  rare pixels
 that lie on the orthogonal complement of the signal subspace, denoted as $\gamma$, for MPSNR =  30.71 dB. }
\label{fig:defineDifficulty}
\end{figure}

A spin-off of the RhyDe denoiser, an anomaly detector is proposed in (\ref{eq:detector}), and classifies pixels outside signal subspace as anomalies.
Therefore, its detection ability depends on the relative power of rare pixels that lie orthogonal to the signal subspace, denoted as $\gamma = \frac{\| ({\bf I} - {\bf EE}^T)({\bf Y}-{\bf N})\|_F}{ \| ({\bf I} - {\bf EE}^T) {\bf N}\|_F}$, where the columns of $\widetilde{\bf Y}$ and $\widetilde{\bf N}$ are noisy rare pixels and noise, respectively. 
The parameter $\gamma$ can provide  an insight into the difficulty of detection.
This can be observed from our experiments, where we simulated images containing anomalous pixels with different values of $\gamma$. 
To save space, we only report the minimum false alarm rates when the detectors reach 100\% detection rates in Fig. \ref{fig:defineDifficulty}. The detection of the outliers becomes more difficult as $\gamma$ decreases.
Compared with classical detectors and sparse-representation based detectors,   RhyDe  almost uniformly yields the best detection results.

\section{Experiments with real images}

The performance of the denoising and anomaly detection using the proposed RhyDe was evaluated using two real HSIs: a SpecTIR image and a sub-scence of the Pavia Centre data. These data contained unknown targets that were spectrally distinct from the image backgrounds. Denoising and anomaly detection were carried out using the same methods as those described in Section~\ref{sec:simexp}.    

\subsection{SpecTIR Data}

\begin{figure}[htbp]
\centering
\includegraphics[width=8cm]{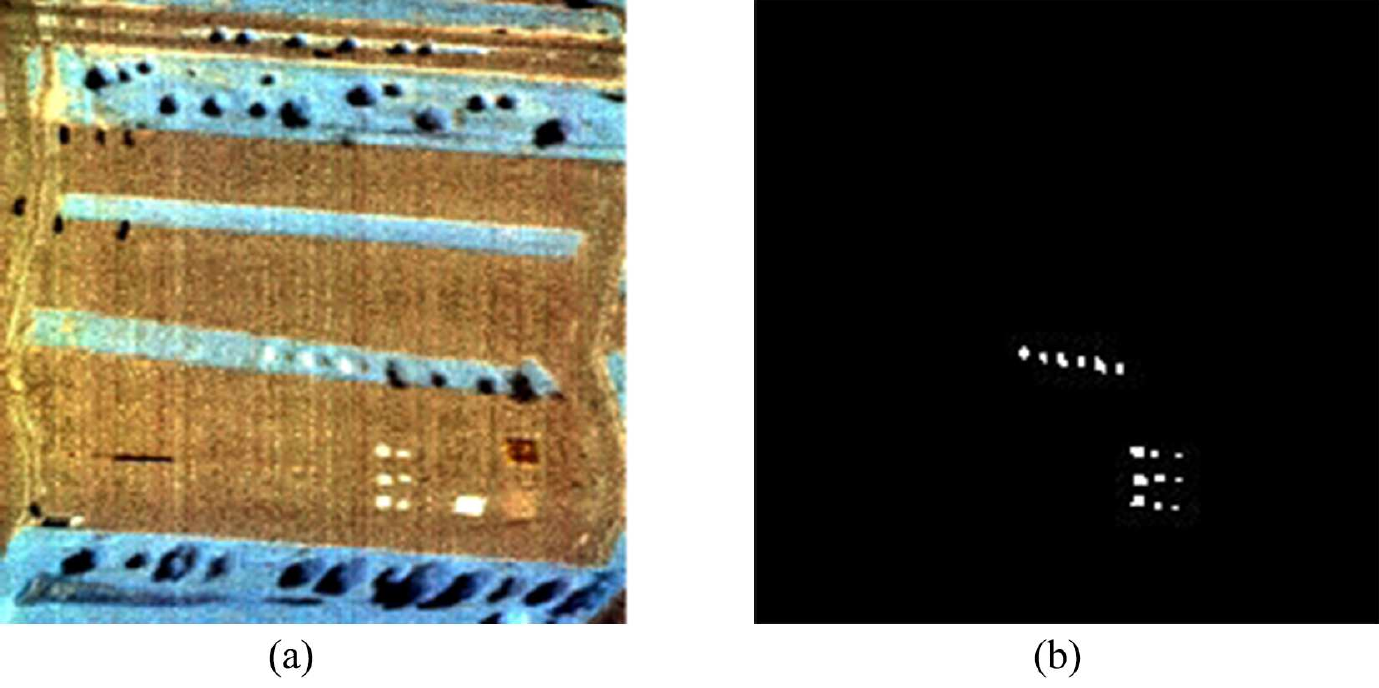}
\caption{(a) Hyperspectral image: specTIR data and (b) ground-truth map of anomalous targets.}
\label{specTIR_img_groundTtri}
\end{figure}

     \begin{figure}[htbp]
\centering
\includegraphics[width=8cm]{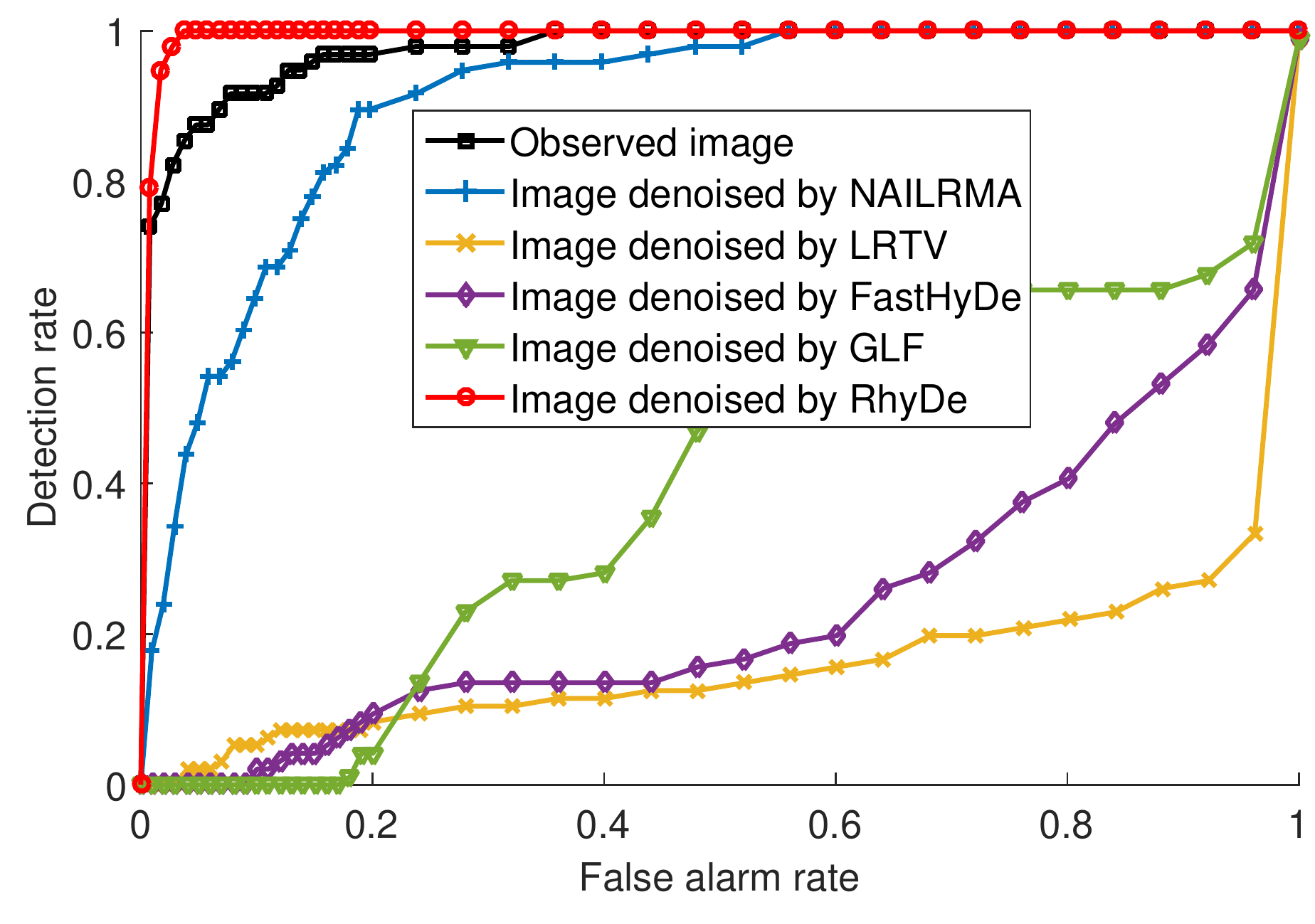}
\caption{ROCs of Global RX detector applying to original and denoised images of specTIR data.}
\label{specTIR_GRXafterDifDenoiser}
\end{figure}

      \begin{figure}[htbp]
\centering
\includegraphics[width=8cm]{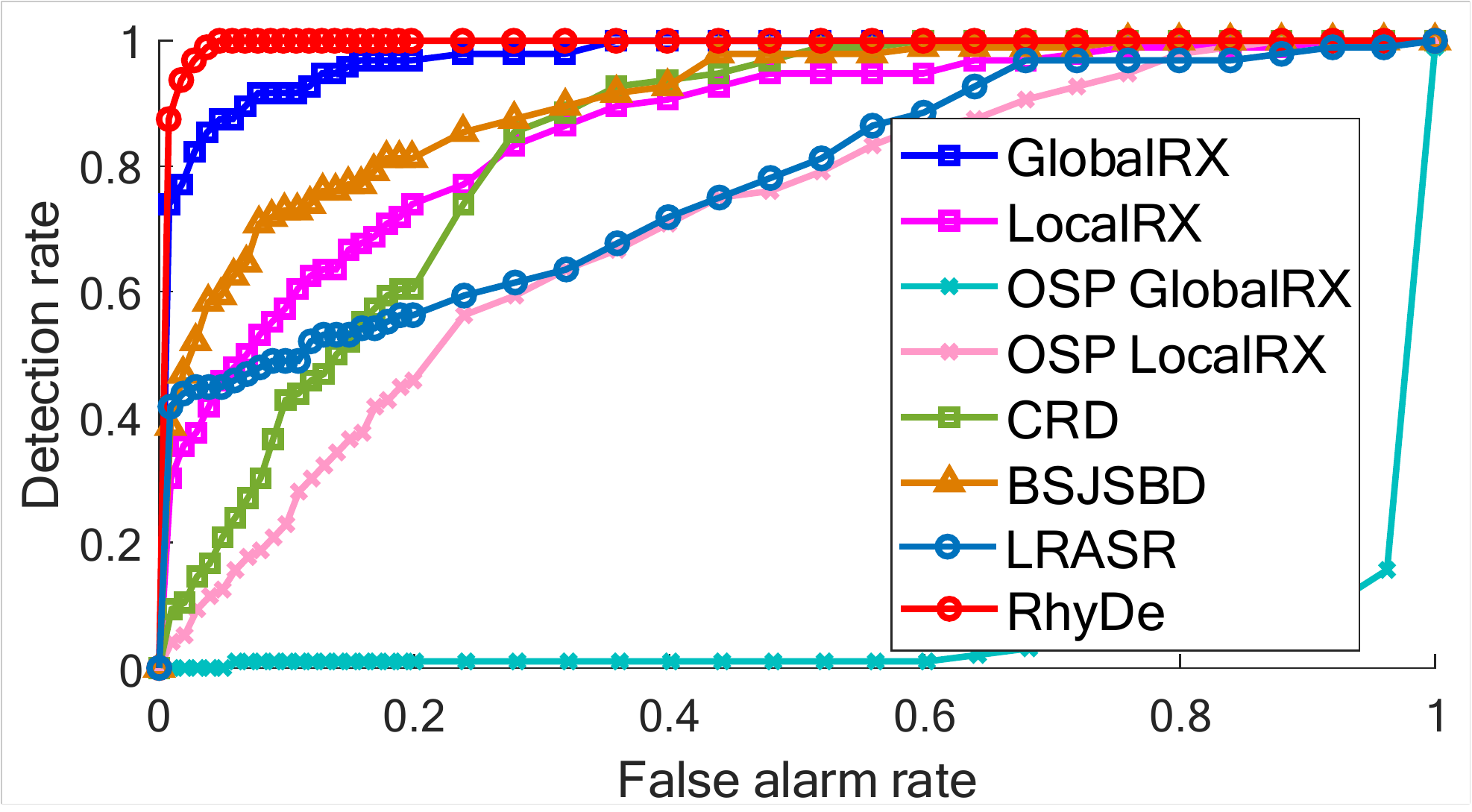}
\caption{ROCs of different anomaly detectors in specTIR data}
\label{spectir_roc}
\end{figure}

    \begin{figure*}[htbp]
\centering
\includegraphics[width=17cm]{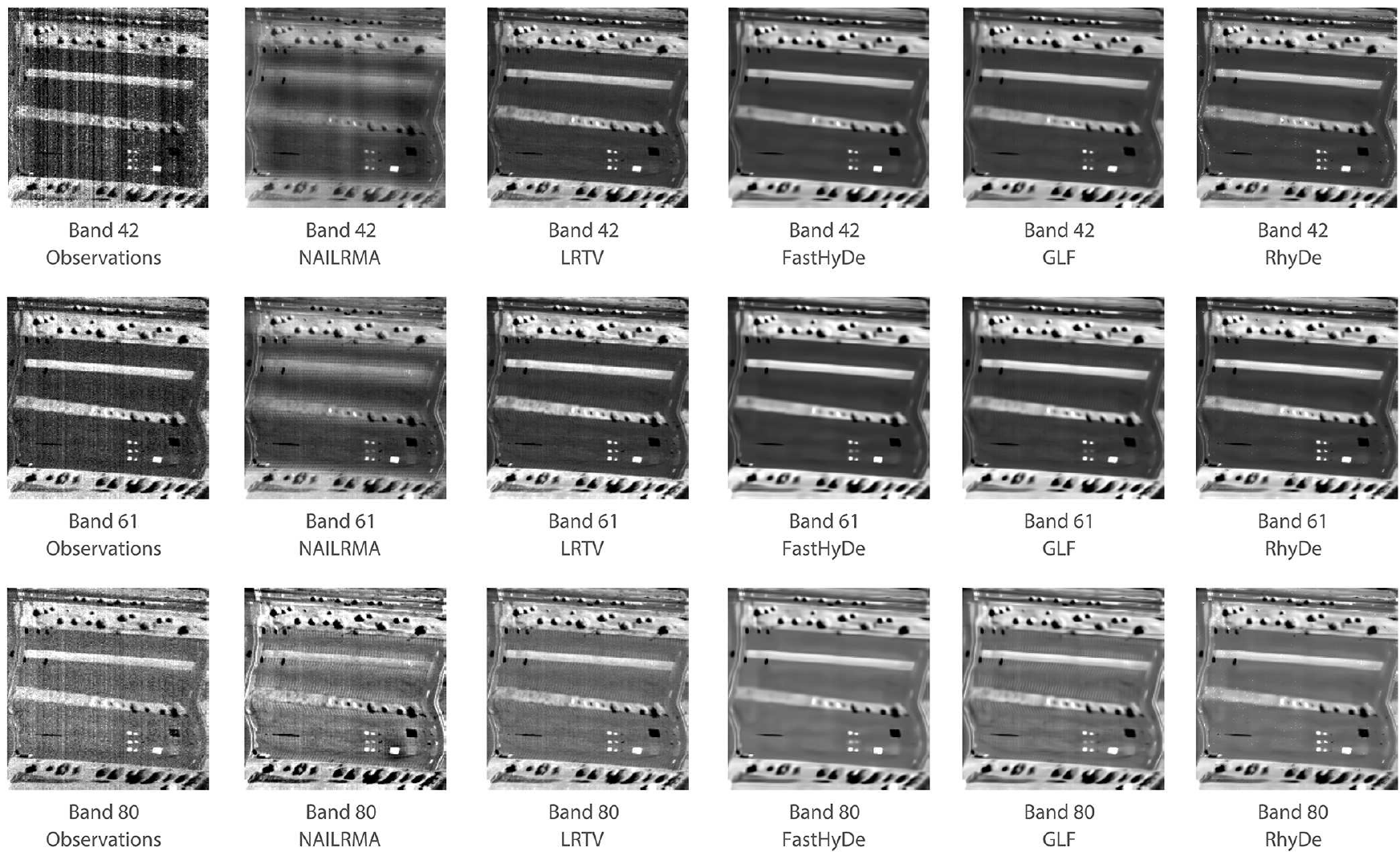}
\caption{Observed Bands 42, 61, and 80 of specTIR data and their denoising results.}
\label{SingleImg_specTIR}
\end{figure*}

 \begin{figure*}[htbp]
\centering
\includegraphics[width=18cm]{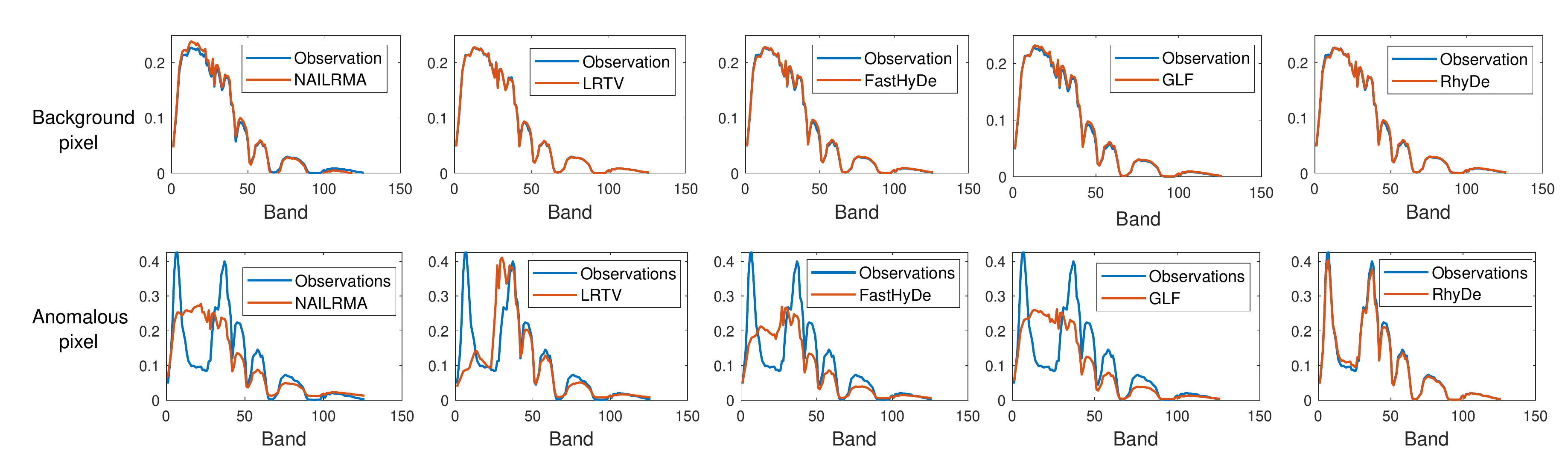}
\caption{Denoised spectral signatures of a background pixel (in the first row)
and an anomalous pixel (in the second row) in specTIR data.}
\label{spec_normal_ano_specTIR}
\end{figure*}

The first real hyperspectral image was acquired by a data collection campaign called the SpecTIR hyperspectral airborne Rochester experiment (SHARE) \cite{herweg2012spectir} on July 29, 2010 using a  ProSpecTIR-VS2 sensor. We extracted a subset of these data to use as a test image. This image is shown in Fig. \ref{specTIR_img_groundTtri}-(a)and consists of 126 bands in the range 390 nm to 2450 nm with a spatial resolution of approximately 1 m. There is a high level of noise in a number of the bands. Fig. \ref{SingleImg_specTIR}  shows three of the bands, namely bands 42, 61, and 80.
As presented in Fig. \ref{specTIR_img_groundTtri}-(a), road and vegetation are the main background materials. Several red and blue fabrics with sizes of 9, 4, and 0.25 m$^2$ were considered to be anomalous targets: ground-truth map for these is shown as Fig. \ref{specTIR_img_groundTtri}-(b). 
These data have also been widely used in other studies  to evaluate the performance of anomaly detectors  \cite{guo2014weighted,gao2014probabilistic}.  

We applied denoising to the SpecTIR data under the assumption of non-i.i.d. noise; however, none of the denoisers worked as well as expected. We supposed that the original hyperspectral data were mainly affected by Poission noise. We applied the Anscombe transform \cite{AnscT}, ${\bf Y} \leftarrow 2\sqrt{{\bf Y} + \frac{3}{8}}$, which can convert  Poisson noise into approximately additive noise. Data that included band-variant noise were whitened before the denoisers were applied \cite{FastHyDe}. 
%
The signal subspace dimension was empirically set to 3 for LRTV, FastHyDe, GLF, and RhyDe. The parameters for NAILRMA and LRTV were hand-tuned to achieve optimal performance.

We compared the visual quality of the denoising results owing to the lack of ground-truth for the noise-free imagery.  As shown by Fig. \ref{SingleImg_specTIR}, LRTV, FastHyDe, GLF, and RhyDe are able to mitigate the effect of the noise in bands 42, 61, and 80, whereas visible noise still remains after applying NAILRMA.

In the case where rare pixels were present, tests were made of whether the denoising would cause the information contained in these pixels to be lost. Fig. \ref{spec_normal_ano_specTIR} shows the denoising results for a background pixel and an anomalous pixel. RhyDe is able to preserve the anomalous pixel, which is an advantage over its low-rank-based competitors. We also applied the classical anomaly detector, Global RX, to the original and denoised images. The detection results, as measured by the ROC, are shown in Fig. \ref{specTIR_GRXafterDifDenoiser}. It can be seen that the detectability of the anomalies is degraded by the low-rank-based denoisers NAILRMA, LRTV, GLF, and FastHyDe but is improved by RhyDe.
 
The detector derived from RhyDe was compared with different detectors in terms of ROC. As shown in Fig. \ref{spectir_roc}, RhyDe performs the best, which implies that relatively complex background land cover can be well represented by a low-rank subspace and that anomalous pixels are preserved in the denoising results obtained using RhyDe.

\subsection{Pavia Centre Sub-scene}

    \begin{figure}[htbp]
\centering
\includegraphics[width=8cm]{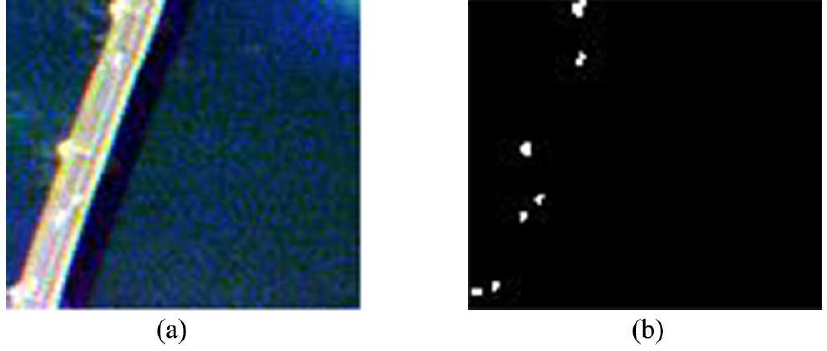}
\caption{(a) Pavia Centre sub-scene and (b) Ground-truth map of vehicles.}
\label{paviaCar_img_groundTtri}
\end{figure}

{\centering
    \begin{figure*}[htbp]

\includegraphics[width=17cm]{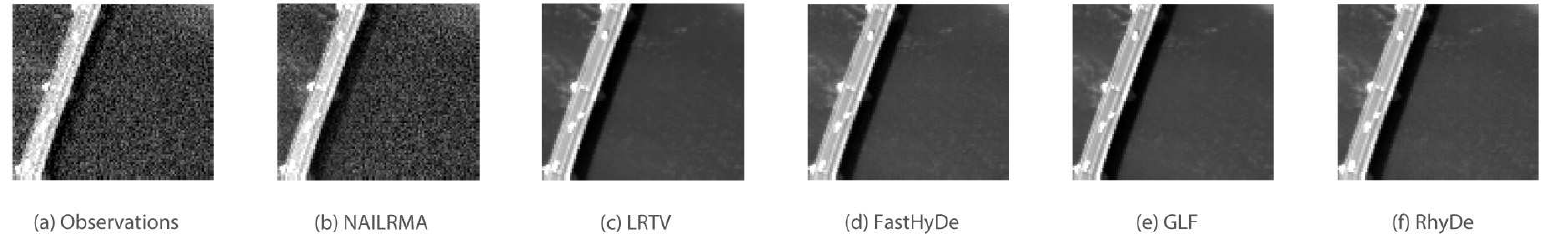}
\caption{Observed Band 6 of Pavia Centre sub-scene and its denoising results.}
\label{SingleImg_paviaCar}

\end{figure*}
 }

 \begin{figure*}[htbp]
\centering
\includegraphics[width=18cm]{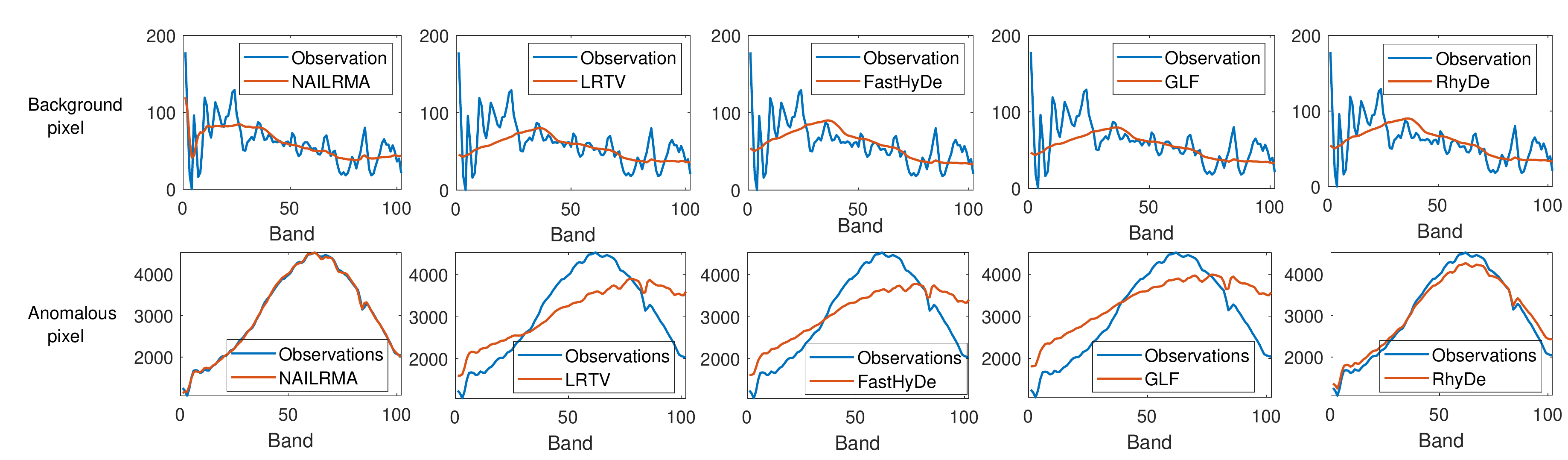}
\caption{Denoised spectral signatures of a background (water) pixel (in the first row)
and an anomalous (vehicle) pixel (in the second row) in Pavia Centre data.}
\label{spec_normal_ano_paviaCar}
\end{figure*}

    \begin{figure}[htbp]
\centering
\includegraphics[width=8cm]{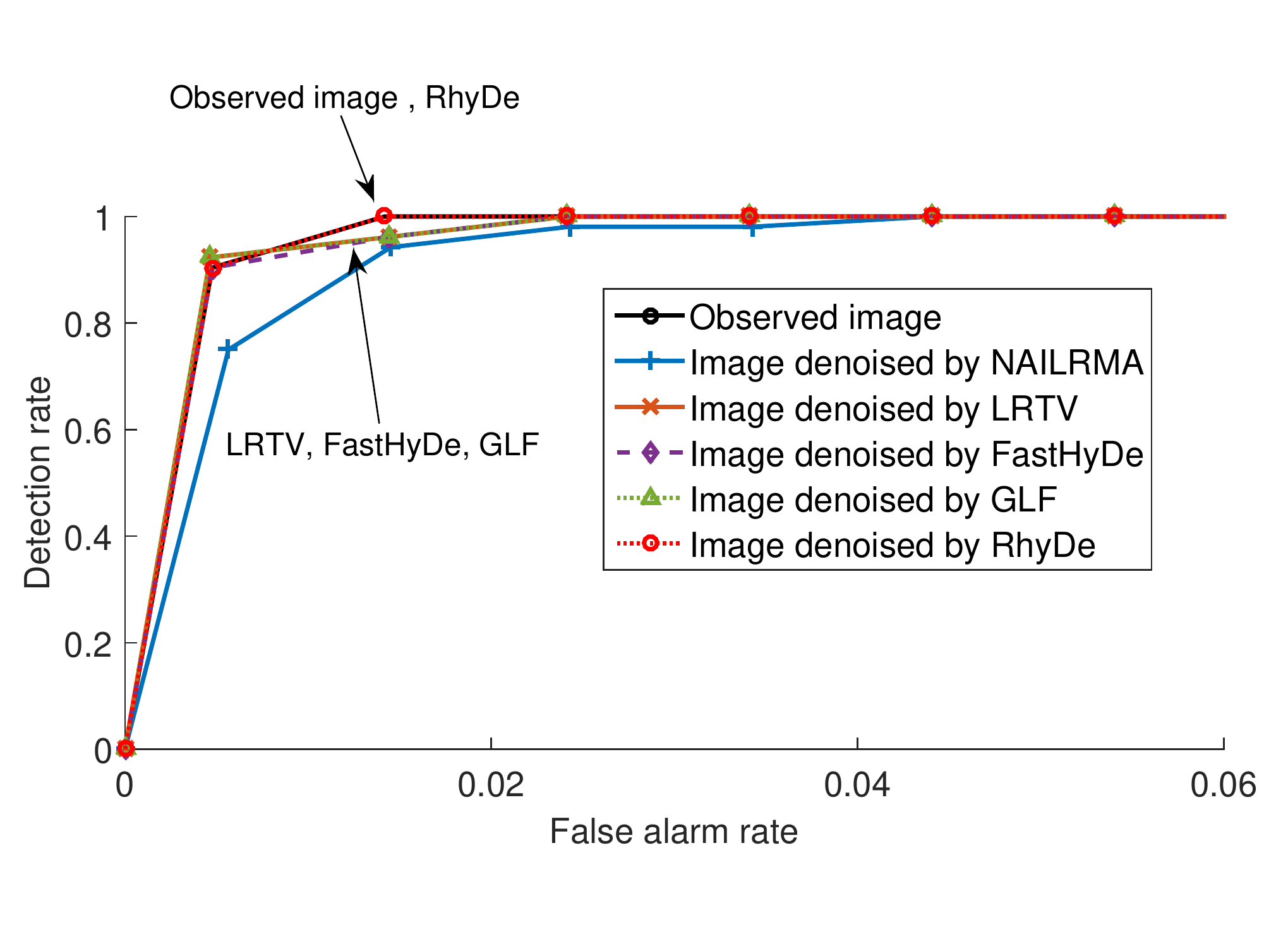}
\caption{ROCs of Global RX detector applying to original and denoised images of Pavia Centre sub-scene.}
\label{GRXafterDifDenoiser_paviaCar}
\end{figure}

    \begin{figure}[htbp]
\centering
\includegraphics[width=8.5cm]{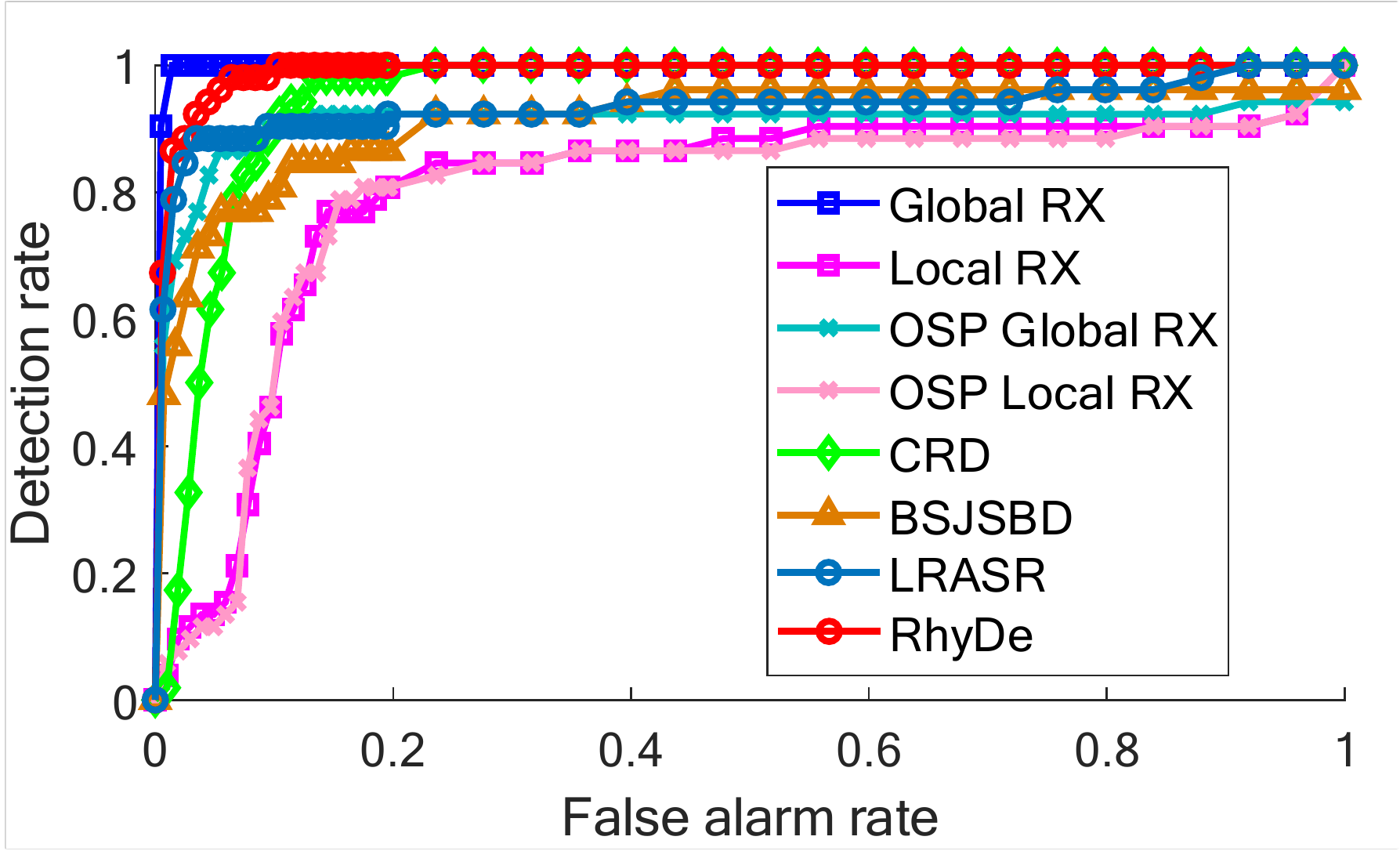}
\caption{ROCs of different anomaly detectors in Pavia Centre sub-scene.}
\label{paviaCar_ROC}
\end{figure}

The second test image had a of size $89 \times 102 \times 102$ (Fig. \ref{paviaCar_img_groundTtri}-(a)). It was a sub-scene of the Pavia Centre data acquired by the ROSIS sensor during a flight campaign over Pavia, northern Italy. The imagery has 102 spectral bands and a spatial resolution of 1.3 m. This sub-scene contained three simple object types: water, a bridge, and vehicles. Suppose that the vehicles are anomalous targets to be detected. A ground-truth map for the vehicles is shown in Fig. \ref{paviaCar_img_groundTtri}-(b). We applied denoisers to the imagery under the assumption of non-i.i.d. noise. The signal subspace dimension was   empirically set to 2 for LRTV, FastHyDe, GLF, and RhyDe. The parameters for NAILRMA and LRTV were hand-tuned to achieve optimal performance.

To demonstrate the denoising performance, because of the lack of ground truth for the noise-free data, the visual quality of the reconstructed bands and reconstructed spectra were used. As shown in Fig. \ref{SingleImg_paviaCar},  the denoisers LRTV, FastHyDe, GLF, and RhyDe clearly produced a great reduction in the amount of noise in band 6. The quality of the reconstructed spectra can be inferred from  Fig. \ref{spec_normal_ano_paviaCar}: here, the first row shows the results for a background pixel (water) and the second row the results for an anomalous pixel (vehicle). It can be seen from Fig. \ref{spec_normal_ano_paviaCar}-(a) that NAILRMA does not remove the noise completely. In addition, only NAILRMA and RhyDe are able to preserve the anomalous spectra in Fig. \ref{spec_normal_ano_paviaCar}-(f, j).

In order to further investigate the impact of the denoising step on the detection of rare pixels, we applied the Global RX detector to the original and denoised images: the detection performance in terms of the ROC is shown in Fig. \ref{GRXafterDifDenoiser_paviaCar}. Only RhyDe does not degrade the detection performance. The anomalous targets (vehicles) in the images denoised by LRTV, FastHyDe, GLF, and NAILRMA are detected with higher false alarm rates than when using the original noisy image, which implies that the spectral signatures of the vehicles are damaged during denoising.

Regarding the anomaly detection, the results given in Fig. \ref{paviaCar_ROC} show that   Global RX performed best with the data used. This is because the image background (namely, water and bridges) in the Pavia Centre sub-scene is simple and can be represented well by the Gaussian density model used in Global RX. Note that RhyDe and CRD also manage to detect all of the targets with a relatively low false alarm rate.

\section{Conclusion}
 
We have proposed RhyDe, a new low-rank-based denoising method, with the aim of preserving rare pixels. As an extension of FastHyDe, the proposed method exploits 
the low-rankness and self-similarity of clean HSIs and the column-wise sparsity of the outlier matrix. A comparison with state-of-the-art denoising algorithms was conducted, and it was concluded that RhyDe yields a better performance with additive noise in terms of the preservation of rare pixels.
The  characteristic of preservation of rare pixels put RhyDe in a privileged position to be used as a pre-procssing step to improve the quality of HSIs, especially, in   situations that rare pixels might be important targets in the subsequent applications, such as anomaly detection and  land-cover change detection. 
 Compared with the other state-of-the-art anomaly detectors, the proposed detector has a comparable detection performance.

 {\small
 
 \bibliographystyle{IEEEbib}
\bibliography{IEEEabrv,summary}
 }
\begin{IEEEbiography}
[{\includegraphics[width=1in,height=1.25in,clip,keepaspectratio]{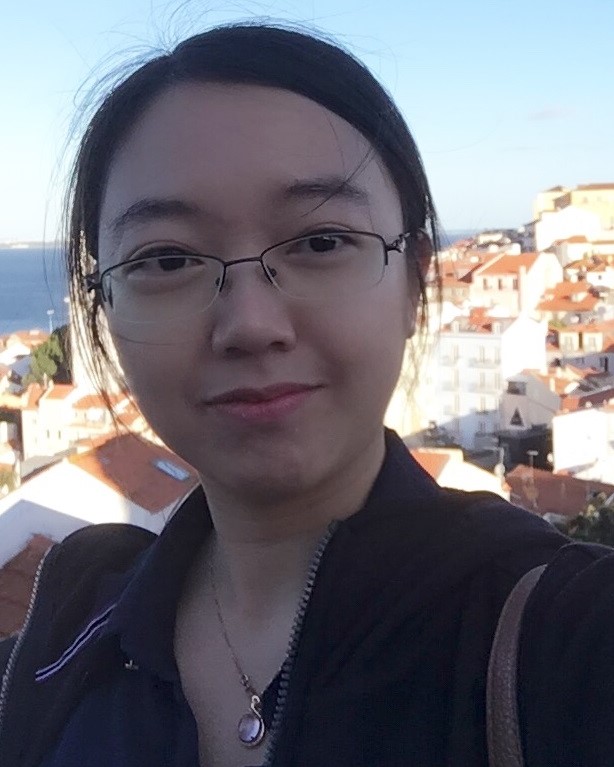}}]{Lina Zhuang}
(S'15-M'20) received Bachelor's  degrees in geographic information system and in economics from South China Normal University, Ghuangzhou, China,  in  2012,  the  M.S.  degree  in  cartography  and  geography  information  system  from  Institute  of  Remote  Sensing  and  Digital  Earth,  Chinese Academy of Sciences, Beijing, China, in 2015, and the Ph.D. degree in Electrical and Computer Engineering at the Instituto Superior Tecnico,  Universidade de Lisboa, Lisbon, Portugal in 2018. 

Since 2015, she has been  a Marie Curie Early Stage Researcher of Sparse Representations and Compressed Sensing Training Network (SpaRTaN number 607290) with the Instituto de Telecomunica\c{c}\~{o}es. SpaRTaN Initial Training Networks (ITN) is funded under the European Union's Seventh Framework Programme  (FP7-PEOPLE-2013-ITN)  call  and  is  part  of  the  Marie  Curie Actions-ITN  funding  scheme.  She is currently  a Research Assistant Professor with Hong Kong Baptist University. Her research interests include hyperspectral image restoration, superresolution, and compressive sensing.
\end{IEEEbiography}

\begin{IEEEbiography}
[{\includegraphics[width=1in,height=1.25in,clip,keepaspectratio]{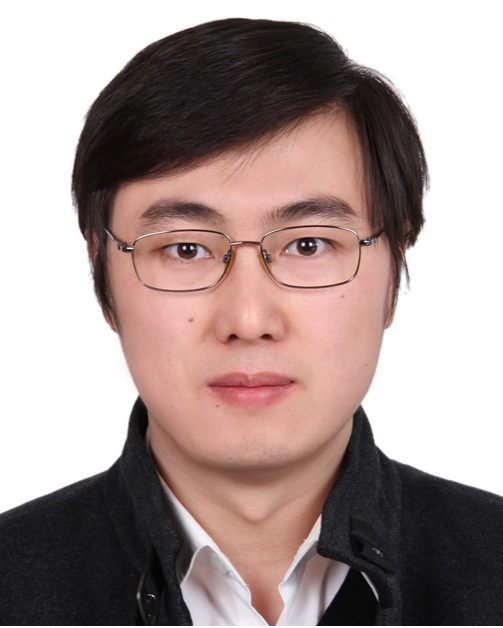}}]{Lianru Gao}
(Senior Member, IEEE) received the B.S. degree in civil engineering from Tsinghua University, Beijing, China, in 2002, and the Ph.D. degree in cartography and geographic information
system from the Institute of Remote Sensing
Applications, Chinese Academy of Sciences (CAS),
Beijing, in 2007.

He is a Professor with the Key Laboratory of Digital
Earth Science, Aerospace Information Research
Institute, CAS. He also has been a Visiting Scholar
with the University of Extremadura, C\'aceres, Spain,
in 2014, and Mississippi State University (MSU), Starkville, MS, USA, in 2016. In last 10 years, he was the PI of ten scientific research projects at national and ministerial levels, including projects by the National Natural Science Foundation of China (2010-2012, 2016-2019, 2018-2020), and by the Key Research Program of the CAS (2013-2015) et al. He has authored or coauthored more than 160 peer-reviewed articles, and there are more than 80 journal articles included by Science Citation Index (SCI). He was a coauthor of an academic book \textit{Hyperspectral Image Classification and Target
Detection}. He holds 28 National Invention Patents in China. His research focuses on hyperspectral image processing and information extraction. Dr. Gao was awarded the Outstanding Science and Technology Achievement Prize of the CAS in 2016, and was supported by the China National Science Fund for Excellent Young Scholars in 2017, and won the Second Prize of The State Scientific and Technological Progress Award in 2018.
He received the recognition of the Best Reviewers of the IEEE JOURNAL OF SELECTED TOPICS IN APPLIED EARTH OBSERVATIONS AND REMOTE SENSING in 2015, and the Best Reviewers of the IEEE TRANSACTIONS ON GEOSCIENCE AND REMOTE SENSING in 2017.
\end{IEEEbiography}

\begin{IEEEbiography}
[{\includegraphics[width=1in,height=1.25in,clip,keepaspectratio]{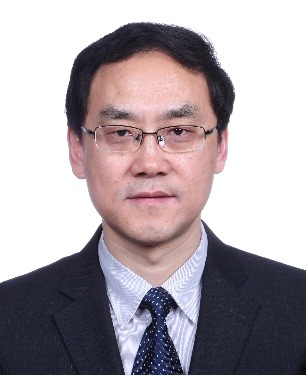}}]{Bing Zhang}
(Fellow, IEEE) received the B.S.
degree in geography from Peking University, Beijing, China, in 1991, and the M.S. and Ph.D. degrees in remote sensing from the Institute of Remote Sensing Applications, Chinese Academy
of Sciences (CAS), Beijing, in 1994 and 2003, respectively.

He is a Full Professor and the Deputy Director of the Aerospace Information Research Institute, CAS, where he has been leading lots of key scientific projects in the area of hyperspectral remote sensing for more than 25 years. He has developed five software systems in the image processing and applications. His creative achievements were rewarded ten important prizes from the Chinese government, and special government allowances of the Chinese State Council. He has authored more than 300 publications, including more than 170 journal articles. He has edited six books/contributed book chapters on hyperspectral image processing and subsequent applications. His research interests include the development of Mathematical and Physical models and image processing software for the analysis of hyperspectral remote sensing data in many different areas.

Dr. Zhang has been serving as a Technical Committee Member of IEEE Workshop on Hyperspectral Image and Signal Processing, since 2011, and has been the President of Hyperspectral Remote Sensing Committee of China National Committee of International Society for Digital Earth since 2012, and has been the Standing Director of Chinese Society of Space Research (CSSR) since 2016. He is the Student Paper Competition Committee Member in IGARSS from 2015 to 2019. He was awarded the National Science Foundation for Distinguished Young Scholars of China in 2013, and the 2016 Outstanding Science and Technology Achievement Prize of the Chinese Academy of Sciences, the highest level of Awards for the CAS scholars. He is serving as an Associate Editor for the IEEE JOURNAL OF SELECTED TOPICS IN APPLIED EARTH OBSERVATIONS AND REMOTE SENSING.
\end{IEEEbiography}

\begin{IEEEbiography}
[{\includegraphics[width=1in,height=1.25in,clip,keepaspectratio]{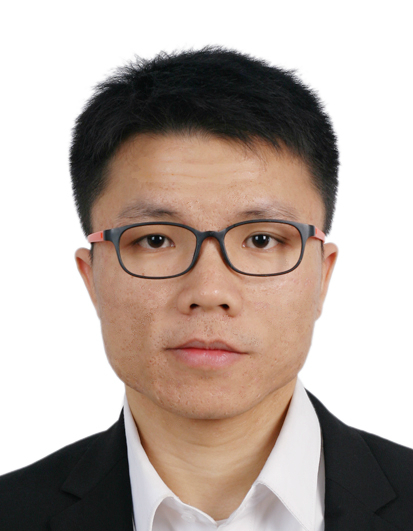}}]{Xiyou Fu}
 received Bachelor's degrees in remote sensing science and technology from Wuhan University, Wuhan, China, in 2012, the M.S. degree in electronic and communication engineering and the Ph.D. degree in cartography and geography information system from Institute of Remote Sensing and Digital Earth, Chinese Academy of Sciences, Beijing, China, in 2015 and 2019, respectively. He is currently a postdoctor with Shenzhen University. His research interests include hyperspectral image restoration, anomaly detection, and superresolution.
\end{IEEEbiography}

 \begin{IEEEbiography}[{\includegraphics[width=1in,height=1.25in,clip,keepaspectratio]{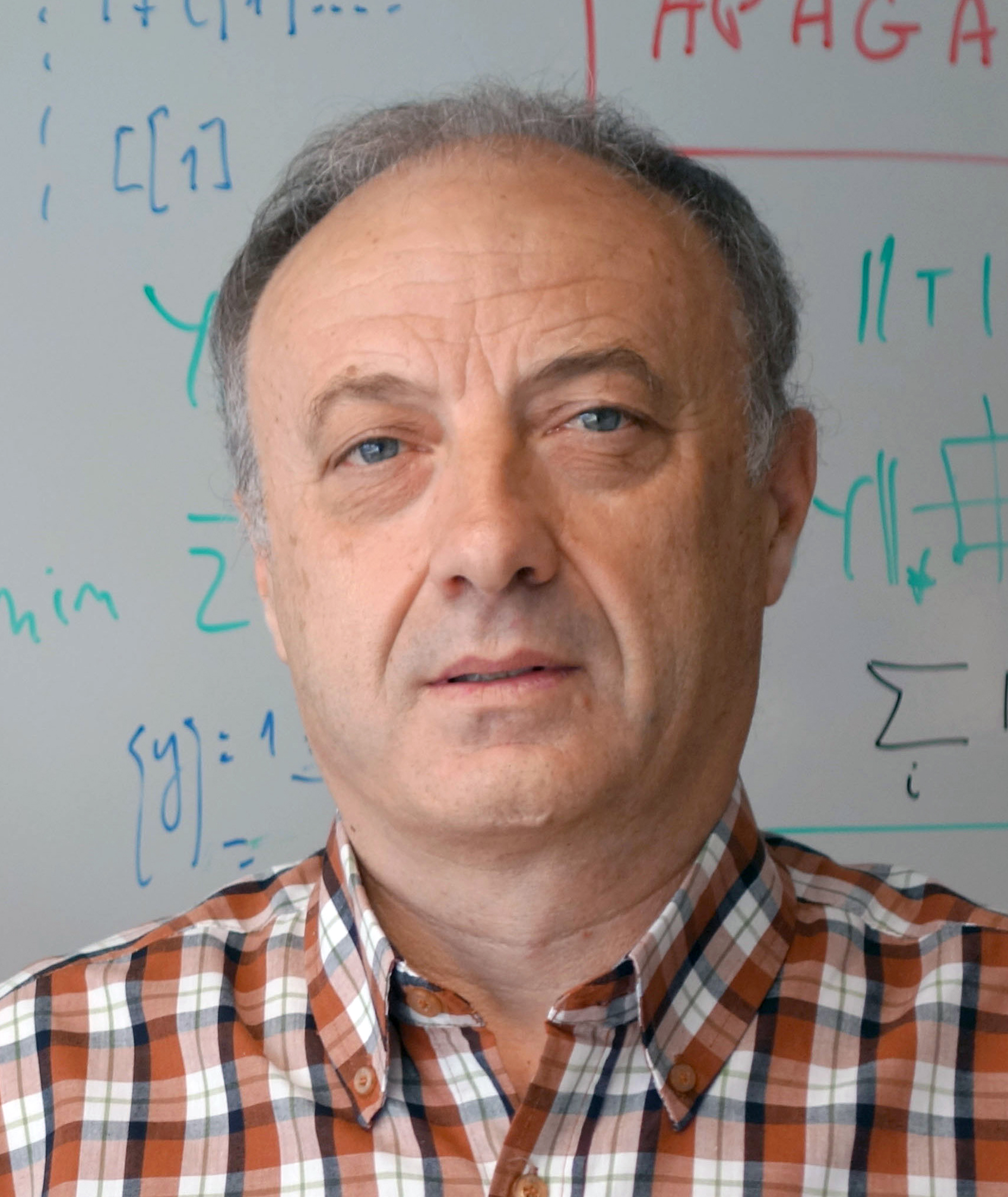}}]{Jos\'{e} M. Bioucas-Dias}
 (S'87--M'95--SM'15--F'17)
received the EE, MSc, PhD, and Habilitation degrees in electrical and
computer engineering from Instituto Superior T\'ecnico (IST),
Universidade T\'ecnica de Lisboa (now Universidade de Lisboa),
Portugal, in 1985, 1991, 1995, and 2007, respectively. Since 1995, he
had been with the Department of Electrical and Computer Engineering,
IST, where he was a Professor and teaches inverse problems
in imaging and electric communications. He was also a Senior Researcher
with the Pattern and Image Analysis group of the Instituto de Telecomunica\c{c}\~{o}es, which is a private non-profit research institution.

His research interests included inverse problems, signal and image
processing, pattern recognition, optimization, and remote sensing. 
He had introduced scientific contributions in the areas  of 
imaging inverse problems, statistical image processing, optimization, 
phase estimation, phase unwrapping, and in various imaging applications, 
such as hyperspectral and radar imaging.  He was included in Thomson
Reuters' Highly Cited Researchers 2015 and 2018 lists and received the 
IEEE GRSS David Landgrebe Award for 2017.
\end{IEEEbiography}
\end{document}